\documentclass[twocolumn]{aa}
\usepackage{graphicx}
\usepackage{txfonts}
\newcounter{subtable}
\def\etal{{et al. \rm}}
\begin{document}
   \title{The photospheric abundances of active binaries}

   \subtitle{III. Abundance peculiarities at high activity level\thanks{Based on observations collected at the European Southern Observatory, Chile (Proposals 64.L-0249 and 071.D-0260).}\fnmsep\thanks{Table~\ref{tab_a1} is only available in electronic form at {\tt http://www.edpsciences.org/aa}}}

   \titlerunning{The photospheric abundances of active binaries III.}

   \author{T. Morel
          \inst{1}
          \and
          G. Micela
          \inst{1}
          \and
          F. Favata
          \inst{2}
          \and
          D. Katz
          \inst{3}
}

   \offprints{T. Morel,
 \email{morel@astropa.unipa.it}}

   \institute{Istituto Nazionale di Astrofisica, Osservatorio Astronomico di Palermo G.\,S. Vaiana, Piazza del Parlamento 1, I-90134 Palermo, Italy
         \and
             Astrophysics Division - Research and Science Support Department of ESA, ESTEC, Postbus 299, NL-2200 AG Noordwijk, The Netherlands
         \and
             Observatoire de Paris, GEPI, Place Jules Janssen, F-92195 Meudon, France
}

   \date{Received 07 April 2004; accepted 11 June 2004}

   \abstract{We report the determination from high-resolution spectra of the atmospheric parameters and abundances of 13 chemical species (among which lithium) in 8 single-lined active binaries. These data are combined with our previous results for 6 other RS CVn systems to examine a possible relationship between the photospheric abundance patterns and the stellar activity level. The stars analyzed are generally found to exhibit peculiar abundance ratios compared to inactive, galactic disk stars of similar metallicities. We argue that this behaviour is unlikely an artefact of errors in the determination of the atmospheric parameters or non-standard mixing processes along the red giant branch, but diagnoses instead the combined action of various physical processes related to activity. The most promising candidates are cool spot groups covering a very substantial fraction of the stellar photosphere or NLTE effects arising from nonthermal excitation. However, we cannot exclude the possibility that more general shortcomings in our understanding of K-type stars (e.g. inadequacies in the atmospheric models) also play a significant role. Lastly, we call attention to the unreliability of the ($V-R$) and ($V-I$) colour indices as temperature indicators in chromospherically active stars.     

\keywords{Stars:fundamental parameters -- stars:abundances -- stars:individual: RS CVn binaries
               }
   }

   \maketitle
%

\section{Introduction} \label{sect_intro}
The class of RS CVn binaries is loosely defined as being composed of systems made up of two late-type, chromospherically active stars, at least one of which having already evolved off the main sequence (Hall 1976). Because the chromospheric activity is rejuvenated by the spinning up of the stars due to the transfer of orbital angular momentum to the rotational spin, they can sustain high activity levels at advanced ages compared to otherwise similar, single stars. A better understanding of this phenomenon, and in particular of the temporal evolution of the stellar activity level (e.g. the coronal X-ray emission), requires a good handle on their evolutionary status (e.g. Barrado \etal 1994). Precise estimates of the atmospheric parameters and abundances are necessary for a robust determination of the isochrone ages, which are very sensitive in cool stars to the choice of $T_{\rm eff}$, [Fe/H] and [$\alpha$/Fe]. The lithium content can also be used to set stringent constraints on the evolutionary status, but accurate effective temperatures are needed. In virtue of the growing body of evidence for distinct abundance trends of some elements in thin and thick disk stars (e.g. Bensby, Feltzing \& Lundstr\"om 2003), elemental abundances can also complement the kinematical data to help establishing whether these objects belong to young or old disk star populations. 

We have started a project that aims at deriving the photospheric parameters and abundances of several key chemical species in a large sample of active binaries. First results for \object{IS Vir} and \object{V851 Cen} have been presented in Katz \etal (2003, hereafter Paper I). Four additional RS CVn binaries have been analyzed in the second paper of this series (Morel \etal 2003, hereafter Paper II). The most salient results of these studies can be summarized as follows: (a) active binaries are not as iron-deficient as previously claimed, (b) they present a large overabundance of most elements with respect to iron and (c) they exhibit photometric anomalies leading to spuriously low ($V-R$) and ($V-I$) colour temperatures. 

These inferences were, however, somewhat tentative because of the small number of objects studied. We have therefore complemented these data by obtaining spectra for 8 additional single-lined RS CVn binaries. Here we use the whole dataset to draw more robust conclusions regarding the behaviour of the photospheric abundances as a function of the activity level and to address the reliability of LTE abundance analyses at the upper end of the stellar activity distribution. 

\section{Observational material} \label{sect_obs}
The spectra were obtained in July 2003 at the ESO 2.2-m telescope (La Silla, Chile) with the
fiber-fed, cross-dispersed echelle spectrograph FEROS in the object+sky configuration. 
The spectral range covered is 3600--9200 \AA \ (resolving power $R$$\sim$48\,000), hence enabling a {\em simultaneous} coverage of \ion{Ca}{ii} H+K and the spectral lines used as abundance indicators. 

\begin{figure*}
\resizebox{\hsize}{!}
{\rotatebox{0}{\includegraphics{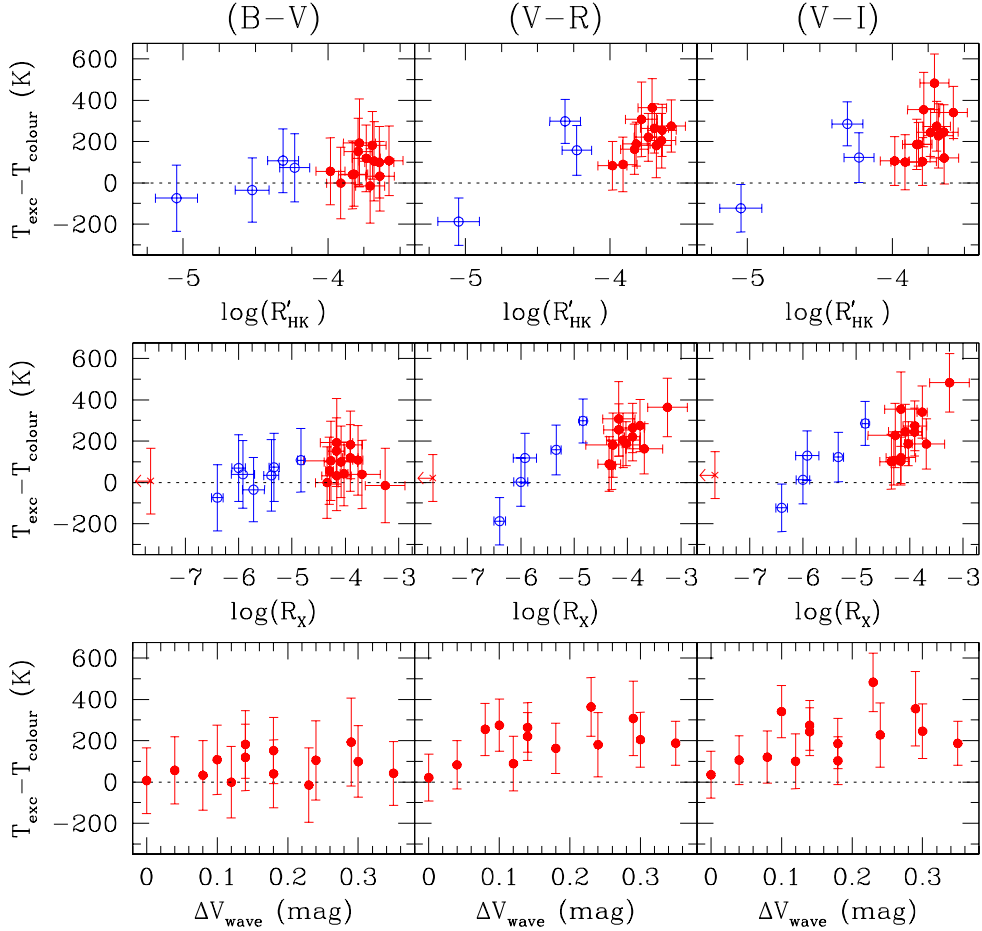}}}
\caption{Differences between the excitation and colour temperatures, as a function of the activity indices $R_{\rm HK}^{\prime}$ ({\em top panels}), $R_{\rm X}$ ({\em middle panels}) and maximum amplitude of the wave-like photometric variations in $V$ band, $\Delta V_{\rm wave}$ ({\em bottom panels}). The variations are shown for temperatures derived from ($B-V$) ({\em left-hand panels}), ($V-R$) ({\em middle panels}) and ($V-I$) data ({\em right-hand panels}). The active binaries and stars in the control sample are indicated by filled and open circles, respectively (in the online version of this journal, red and blue circles refer to active and control stars, respectively). The crosses are upper limits.} 
\label{fig_temperature}
\end{figure*}

The 8 main programme stars (all of spectral type G8--K2 IV--III) are drawn from the catalogue of chromospherically active stars of Strassmeier \etal (1993). Since fast rotators are not amenable to a curve-of-growth abundance analysis because of blending problems, only stars satisfying $v \sin i \la 10$ km s$^{-1}$ were considered. Seven presumably single stars of similar spectral type, but with a much lower level of X-ray emission (H\"unsch, Schmitt \& Voges 1998) were also observed. They were selected on the basis of their colours and absolute magnitudes. Because of observational constraints, the reader should bear in mind that the control stars do not exactly match the properties of the target sample: they are on average slightly bluer and younger. Some basic properties of the targets and further details on the observing campaign (mean heliocentric Julian date of the observations, typical signal-to-noise ratio) can be found in Tables 1$a$-$b$. As will be apparent in the following, one star originally classified as an active binary (\object{HD 28}, \object{33 Psc}) actually displays an extremely low level of activity (see also Walter 1985).  

\begin{figure*}
\resizebox{\hsize}{!}
{\rotatebox{0}{\includegraphics{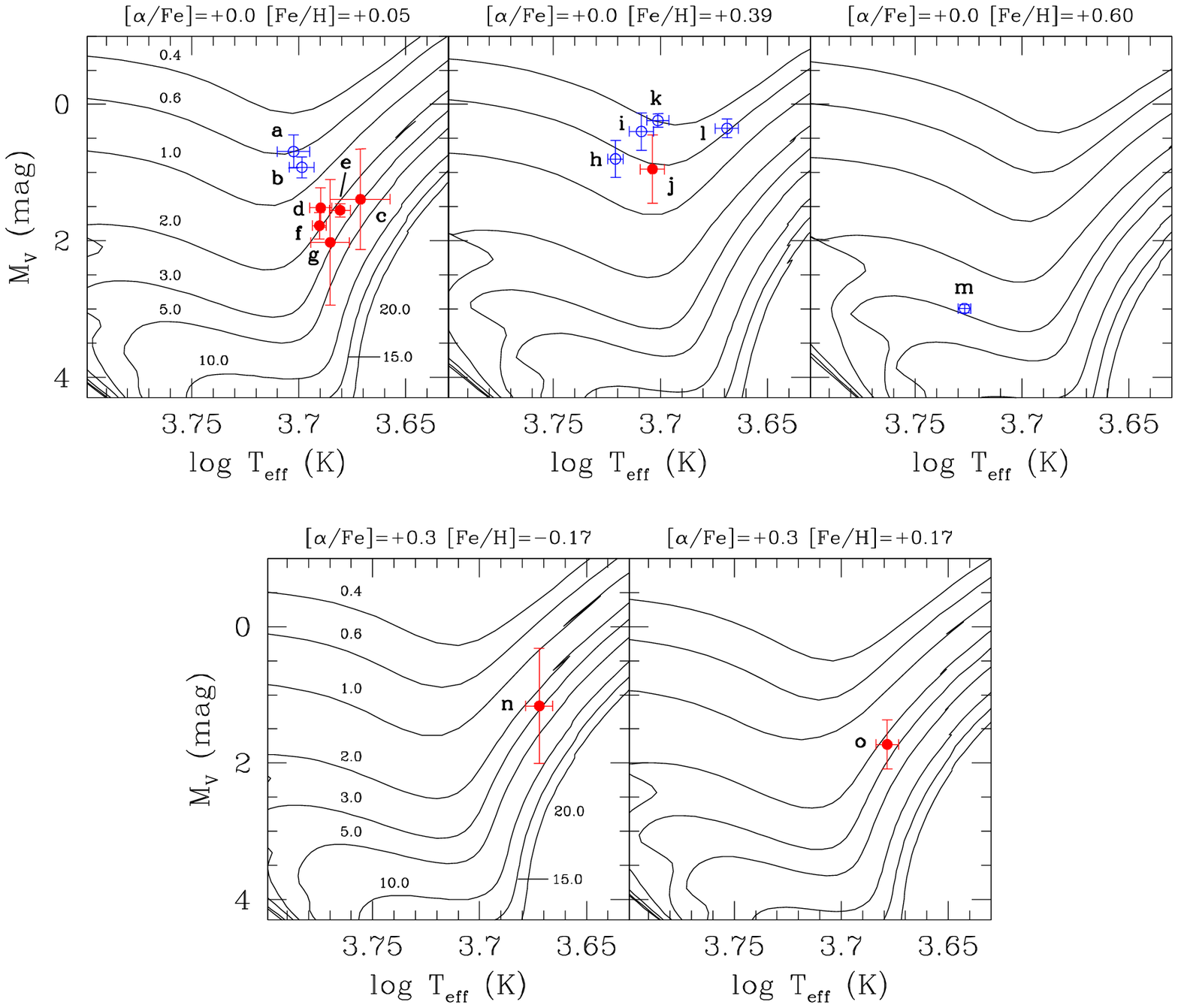}}}
\caption{Positions of the programme stars in HR diagrammes: ({\em a}) \object{HD 218527}, ({\em b}) \object{HD 4482}, ({\em c}) \object{HD 182776}, ({\em d}) \object{HD 217188}, ({\em e}) \object{HD 28}, ({\em f}) \object{HD 181809}, ({\em g}) \object{HD 204128}, ({\em h}) \object{HD 154619}, ({\em i}) \object{HD 1227}, ({\em j}) \object{HD 205249}, ({\em k}) \object{HD 211391}, ({\em l}) \object{HD 156266}, ({\em m}) \object{HD 17006}, ({\em n}) \object{HD 19754} and ({\em o}) \object{HD 202134}. The metallicity of the isochrones refers to a solar iron abundance: $\log \epsilon_{\odot}$(Fe)=7.50 (Yi et al. 2001; Kim et al. 2002). The age of the isochrones (in Gyr) is indicated in the left-hand panels. Symbols as in Fig.~\ref{fig_temperature}.} 
\label{fig_iso}
\end{figure*}

\renewcommand{\thetable}{\arabic{table}\alph{subtable}}
\addtocounter{table}{0}
\setcounter{subtable}{1}
\begin{table*}
\caption{Physical parameters of the active binaries. Spectral type, rotational and orbital periods from Strassmeier \etal (1993). We quote the typical S/N ratio at 6700 \AA \ after combination of the individual exposures.  The $A(V)$ values are derived from the empirical galactic extinction model of Arenou, Grenon \& G{\'o}mez (1992). The colours in the Cousins and Kron photometric systems have been transformed into Johnson values following Bessell (1979). $\Delta V_{\rm wave}$ is the maximum amplitude of the
wave-like photometric variations in $V$ band (Strassmeier \etal 1993). $v_r$ is the radial velocity adopted in the computation of the velocity components. The peculiar space velocity, $S$, is given
by: $S=(U^2+V^2+W^2)^{1/2}$. The last two
  rows give the stellar ages and masses derived from the evolutionary tracks.}
\label{tab_1a}
\hspace*{-0.8cm}
\begin{tabular}{lcccccccc} \hline\hline
 &  \object{HD 28}  & \object{HD 19754}  & \object{HD 181809} &  \object{HD 182776} &  \object{HD 202134}  & \object{HD 204128}  & \object{HD 205249}  & \object{HD 217188}\\
Name & \object{33 Psc} & \object{EL Eri} & \object{V4138 Sgr} & \object{V4139 Sgr} & \object{BN Mic} & \object{BH Ind} & \object{AS Cap} & \object{AZ Psc} \\\hline
Spectral Type & K0 III & G8 IV-III & K1 III & K2-3 III & K1 IIIp & K1 IIICNIVp & K1 III & K0 III \\
$v\sin i$ (km s$^{-1}$) & $\la$1.9$^{[1]}$ & 7.0$^{[1]}$ & 5.1$^{[1]}$ & $\sim$12$^{[2]}$ & $\la$8$^{[2]}$ & & 7.3$^{[1]}$ & 3.0$^{[1]}$\\
$P_{\rm rot}$ (d) & & 47.96 & 60.23 & 45.18 & 61.73 & 22.35 & 57.90 & 91.2\\
$P_{\rm orb}$ (d) & 72.93 & 48.263 & 13.048 & 45.180 & 63.09 & 22.349 & 49.137 & 47.121\\
$HJD$--2,452,800  & 30.86 & 31.93 & 29.72 & 29.83 & 29.86 & 31.87 & 30.71 & 30.77\\
$S/N$ & 180 & 280 & 265 & 190 & 310 & 230 & 205 & 200\\
$A(V)$ (mag) & 0.099 & 0.227 & 0.212 & 0.239 & 0.191 & 0.163 & 0.147 & 0.070\\
$(B-V)_0$ (mag) & 1.010$^{[3]}$ & 1.026$^{[4]}$ & 0.959$^{[4]}$ & 1.161$^{[4]}$ & 1.073$^{[4]}$ & 0.969$^{[4]}$ & 0.983$^{[4]}$ & 1.010$^{[5]}$\\
$(V-R)_0$ (mag) & 0.755$^{[6]}$ & 0.837$^{[4]}$ & 0.774$^{[4]}$ & 0.919$^{[4]}$ & 0.842$^{[4]}$ & 0.874$^{[4]}$ & 0.752$^{[4]}$ & \\
$(V-I)_0$ (mag) & 1.268$^{[6]}$ & 1.423$^{[4]}$ & 1.294$^{[4]}$ & 1.554$^{[4]}$ & 1.415$^{[4]}$ & 1.533$^{[4]}$ & 1.255$^{[4]}$ & 1.249$^{[5]}$ \\
$T_{\rm colour}$ ($B-V$) (K)$^a$  &  4787  & 4661 &  4860  & 4497  & 4651  & 4860  & 4873  & 4743\\
$T_{\rm colour}$ ($V-R$) (K)$^b$  &  4773 &  4537  & 4714  & 4382  & 4549  & 4481  & 4791 &  \\
$T_{\rm colour}$ ($V-I$) (K)$^b$  &  4759  & 4514  & 4715 &  4335 &  4526  & 4362 &  4781  & 4792\\
$T_{\rm exc}$ (K)  &  4794$\pm$54  & 4700$\pm$69 &  4902$\pm$36   &4690$\pm$151  & 4770$\pm$58 &  4845$\pm$101 &  5055$\pm$67  & 4895$\pm$58\\
log $g$ (cm s$^{-2}$)    &  2.86$\pm$0.13   &  2.30$\pm$0.16  &   2.95$\pm$0.10   &  2.53$\pm$0.28  &   2.63$\pm$0.09   &  2.88$\pm$0.20    & 3.00$\pm$0.09   &   2.81$\pm$0.09\\
$\xi$ (km s$^{-1}$)   &    1.21$\pm$0.06   &  1.77$\pm$0.10   &  1.62$\pm$0.05   &  2.43$\pm$0.26   &  1.70$\pm$0.07    & 1.74$\pm$0.14     &1.89$\pm$0.09    &  1.82$\pm$0.07\\
log ($R_{\rm HK}^{\prime}$)$^c$          &           & --3.83$\pm$0.10   & --3.82$\pm$0.10   & --3.78$\pm$0.11   & --3.74$\pm$0.10   & --3.71$\pm$0.10 & --3.69$\pm$0.10   &  --3.79$\pm$0.10     \\
log($L_{\rm X}$) (ergs s$^{-1}$)$^d$     & $<$27.34$^{[7]}$  &   31.43$\pm$0.34$^{[8]}$  &   30.92$\pm$0.07$^{[7]}$  &   30.88$\pm$0.30$^{[9]}$  &   30.99$\pm$0.15$^{[10]}$  &   31.50$\pm$0.37$^{[10]}$ & 31.27$\pm$0.22$^{[8]}$  &  30.69$\pm$0.16$^{[8]}$    \\
log ($R_{\rm X}$)$^d$                    & $<$--7.65 &   --3.69$\pm$0.34 &   --4.03$\pm$0.07 &   --4.16$\pm$0.30 &   --3.91$\pm$0.15 &   --3.25$\pm$0.37 & --3.90$\pm$0.22 &  --4.17$\pm$0.16    \\
$\Delta V_{\rm wave}$ (mag)    &  0.00   & 0.18   & 0.35  &  0.29   & 0.14  &  0.23   & 0.14  &   0.18\\
$v_r$ (km s$^{-1}$) & --8.0$\pm$4.8$^{[1]}$ & +13.7$\pm$2.0$^{[11]}$ & --12.7$\pm$2.0$^{[11]}$ & --39.1$\pm$2.0$^{[11]}$ & +49.1$\pm$2.0$^{[11]}$ &+7.4$\pm$2.0$^{[11]}$ & --27.0$\pm$2.0$^{[11]}$ & --18.5$\pm$1.4$^{[1]}$\\
$U$ (km s$^{-1}$) & --16.9$\pm$0.5 & --51.3$\pm$13.0 & --15.7$\pm$1.9 & --60.5$\pm$5.4 & 57.2$\pm$5.7 & --15.4$\pm$5.4 & --40.6$\pm$3.3 & --53.9$\pm$5.0\\
$V$ (km s$^{-1}$) & 5.25$\pm$2.12 & --41.6$\pm$14.0 & --47.8$\pm$3.0 & --12.7$\pm$3.2 & --33.4$\pm$6.2 & --7.26$\pm$1.27 & --15.8$\pm$1.4 & --20.5$\pm$1.3\\
$W$ (km s$^{-1}$) & 6.44$\pm$4.37 & 12.5$\pm$12.0 & --22.6$\pm$1.6 & --24.3$\pm$11.2 & --16.1$\pm$4.3 & --24.0$\pm$5.2 & 1.64$\pm$2.73 & --4.67$\pm$2.01\\
$S$ (km s$^{-1}$)  &    18.8$\pm$1.7   &  67.2$\pm$13.4   &  55.2$\pm$2.8   &  66.4$\pm$6.4   &  68.2$\pm$5.8   &  29.4$\pm$5.1    & 43.6$\pm$3.1  &    57.8$\pm$4.7\\
$\tau_{\rm iso}$ (Gyr)  &    2.3$^{+0.3}_{-0.3}$  &   3$^{+6}_{-2}$   &  2.0$^{+0.2}_{-0.4}$  &   3.0$^{+4.5}_{-1.6}$    & 2.5$^{+1.0}_{-1.0}$    & 2.9$^{+3.1}_{-1.7}$  &  0.6$^{+0.3}_{-0.1}$   &   1.5$^{+0.5}_{-0.3}$\\
$M$ (M$_{\sun}$) & $\sim$1.7 & $\sim$1.5 & $\sim$1.7 & $\sim$1.5 & $\sim$1.6 & $\sim$1.5 & $\sim$2.6 & $\sim$1.9 \\\hline
\end{tabular}
\end{table*}

\addtocounter{table}{-1}
\addtocounter{subtable}{1}
\begin{table*}
\caption{Physical parameters of the control stars.}
\label{tab_1b}
\hspace*{-0.5cm}
\begin{tabular}{lccccccc} \hline\hline
 &  \object{HD 1227} & \object{HD 4482} & \object{HD 17006}&  \object{HD 154619} & \object{HD 156266} & \object{HD 211391} & \object{HD 218527} \\\hline
Spectral Type & G8 II-III & G8 II & K1 III & G8 III-IV & K2 III & G8 III-IV & G8 III-IV\\
$v\sin i$ (km s$^{-1}$) & $\la$1.0$^{[1]}$ & $\la$1.0$^{[1]}$ & & 1.3$^{[1]}$ & $\la$15$^{[12]}$ & $\la$15$^{[12]}$ & 3.0$^{[13]}$\\
$S/N$ & 190 & 205 & 190 & 190 & 180 & 225 & 115\\
$A(V)$ (mag) & 0.070 & 0.069 & 0.052 & 0.247 & 0.267 & 0.096 & 0.070\\
$(B-V)_0$ (mag) & 0.899$^{[14]}$ & 0.949$^{[15]}$ & 0.864$^{[14]}$ & 0.795$^{[16]}$ & 1.069$^{[14]}$ & 0.961$^{[14]}$ & 0.889$^{[14]}$\\
$(V-R)_0$ (mag) & 0.683$^{[17]}$ & 0.733$^{[17]}$ & 0.677$^{[17]}$ & & 0.734$^{[6]}$ & 0.676$^{[6]}$ & \\
$(V-I)_0$ (mag) & 1.144$^{[17]}$ & 1.204$^{[17]}$ & 1.113$^{[17]}$ & & 1.251$^{[6]}$ & 1.130$^{[6]}$ & \\
$T_{\rm colour}$ ($B-V$) (K)$^a$ &  5076&  4922&  5224 & 5295 & 4739 & 4955&  5006  \\
$T_{\rm colour}$ ($V-R$) (K)$^b$ &  4997  &4837 & 5034 &  & 4853 & 5023 &   \\
$T_{\rm colour}$ ($V-I$) (K)$^b$  & 4985 & 4872 & 5046&   & 4788 & 5012 &   \\
$T_{\rm exc}$ (K)  & 5115$\pm$67 & 4995$\pm$67 & 5332$\pm$37 & 5260$\pm$44 & 4665$\pm$58 &  5025$\pm$60 & 5040$\pm$88 \\ 
log $g$ (cm s$^{-2}$)    &  3.04$\pm$0.10   & 2.88$\pm$0.08  &  3.80$\pm$0.09 &   3.38$\pm$0.09  &  2.50$\pm$0.13  &  2.84$\pm$0.08  &  2.79$\pm$0.20  \\  
$\xi$ (km s$^{-1}$)    & 1.33$\pm$0.06  &  1.47$\pm$0.07  &  1.22$\pm$0.06  &  1.28$\pm$0.07  &  1.48$\pm$0.06  &  1.47$\pm$0.07   & 1.23 $\pm$0.12   \\
log ($R_{\rm HK}^{\prime}$)$^c$          &    & --4.23$\pm$0.10   & --4.31$\pm$0.10   & --4.52$\pm$0.11   &  --5.05$\pm$0.12   &    &    \\
log($L_{\rm X}$) (ergs s$^{-1}$)$^d$     & 29.39$\pm$0.23$^{[18]}$   & 29.83$\pm$0.10$^{[18]}$   & 29.45$\pm$0.07$^{[18]}$   &  29.52$\pm$0.20$^{[18]}$  &  28.97$\pm$0.18$^{[18]}$   & 29.42$\pm$0.13$^{[18]}$   &  29.86$\pm$0.14$^{[18]}$  \\
log ($R_{\rm X}$)$^d$                    &  --5.92$\pm$0.23  & --5.34$\pm$0.10    & --4.83$\pm$0.07   & --5.72$\pm$0.20   &  --6.46$\pm$0.18   & --6.00 $\pm$0.13  &  --5.39$\pm$0.14  \\
$v_r$ (km s$^{-1}$) & --0.3$\pm$0.2$^{[1]}$ & --5.1$\pm$1.5$^{[1]}$ & +13.0$\pm$2.0$^{[19]}$ & --22.8$\pm$0.2$^{[1]}$ & --2.3$\pm$0.9$^{[19]}$ & --14.7$\pm$0.9$^{[19]}$ & --17.8$\pm$2.0$^{[19]}$ \\
$U$ (km s$^{-1}$) & 8.05$\pm$2.56 & --21.0$\pm$1.0 & 2.63$\pm$0.50 & --27.9$\pm$0.4 & --4.06$\pm$0.96 & --40.8$\pm$1.2 & --83.6$\pm$7.8 \\
$V$ (km s$^{-1}$) & --3.13$\pm$0.77 & --27.7$\pm$1.6 & --23.3$\pm$1.2 & --0.90$\pm$1.61 & --27.1$\pm$1.2 & --24.2$\pm$0.9 & --1.96$\pm$2.19 \\
$W$ (km s$^{-1}$) & --9.95$\pm$0.59 & --12.0$\pm$1.3 & --13.0$\pm$1.8 & --33.0$\pm$1.6 & --8.83$\pm$0.56 & --16.7$\pm$1.2 & 6.80$\pm$1.76\\
$S$ (km s$^{-1}$) &   13.2$\pm$1.6  &  36.8$\pm$1.4  &  26.8$\pm$1.3  &  43.2$\pm$1.3   & 28.8$\pm$1.2    &50.3$\pm$1.1   & 83.9$\pm$7.8 \\    
$\tau_{\rm iso}$ (Gyr) &   0.5$^{+0.1}_{-0.1}$ &  0.8$^{+0.1}_{-0.1}$ &   2.9$^{+0.1}_{-0.1}$  & 0.7$^{+0.1}_{-0.1}$ &  0.6$^{+0.1}_{-0.1}$  & 0.40$^{+0.05}_{-0.05}$ &  0.6$^{+0.2}_{-0.1}$ \\   
$M$ (M$_{\sun}$) & $\sim$2.8 & $\sim$2.4 & $\sim$1.4 & $\sim$2.6 & $\sim$2.6 & $\sim$3.0 & $\sim$2.6 \\\hline
\end{tabular}
Key to references --- [1] de Medeiros \& Mayor (1999); [2] Randich, Gratton \& Pallavicini (1993); [3] Walter (1985); [4] Evans \& Koen (1987); [5] {\it Hipparcos} data; [6] Johnson \etal (1966a); [7]  Dempsey \etal (1993); [8] Schwope \etal (2000); [9] Neuh\"auser \etal (2000); [10] Dempsey \etal (1997); [11] Balona (1987); [12] Herbig \& Spalding (1955); [13] Gray \& Nagar (1985); [14] Johnson \etal (1966b); [15] H\"aggkvist \& Oja (1987); [16] Oja (1991); [17] Eggen (1989); [18] H\"unsch \etal (1998); [19] SIMBAD database.\\
$^a$ The typical 1-$\sigma$ uncertainty is 150 K.\\
$^b$ The typical 1-$\sigma$ uncertainty is 100 K.\\
$^c$ A blank indicates that the \ion{Ca}{ii} H+K profiles were not filled in by emission. The errors take into account the uncertainties in $(V-R)_0$ and $T_{\rm eff}$.\\
$^d$ All luminosities were rescaled to the {\em Hipparcos} distances.  The errors take into account the uncertainties in the X-ray counts and in the distances.\\
\end{table*}

Tasks implemented in the {\tt IRAF}\footnote{{\tt IRAF} is distributed by the National Optical Astronomy Observatories, operated by the Association of Universities for Research in Astronomy, Inc., under cooperative agreement with the National Science Foundation.} package were used to carry out the standard reduction procedures for echelle spectra (i.e. bias subtraction, flat-field correction, removal of scattered light, order extraction and wavelength calibration). The continuum normalization was performed using Kurucz synthetic spectra (Kurucz 1993) calculated for an initial guess of the atmospheric parameters and metallicities. When available, temperature and metallicity values were taken from the compilations of Taylor (2003) and Cayrel de Strobel, Soubiran \& Ralite (2001), respectively. These preliminary temperature estimates differ by less than 125 K from our final values derived from the excitation equilibrium of the iron lines. Otherwise, the temperatures were determined from ($B-V$) indices (Alonso, Arribas \& Mart\'{\i}nez-Roger 1999, 2001) or, in a few cases, calibrated on the spectral type (Flower 1996). If unknown, solar metallicity was assumed. We adopted for all the stars $\log g$=2.7 cm s$^{-2}$. In the vast majority of cases, this estimate is in resonable agreement both with previous works (Cayrel de Strobel \etal 2001) and with our final values. The $v \sin i$ values used are quoted in Tables 1$a$-$b$. The line-free regions were selected from these synthetic spectra and then fitted by low-degree polynomials in the observed spectra. Every spectral line in our list was inspected in the 3 consecutive exposures obtained in order to identify the spectra with an imperfect continuum normalization or cosmic ray events. Any individual exposure affected by such problems was excluded during the co-adding procedure. Finally, we corrected for the star's systemic velocity by Doppler-shifting the spectra by the mean radial velocity of the lines under study.

\section{Analysis} \label{sect_analysis}
To ensure consistency, the abundance analysis carried out here is identical in all respects to the procedure employed in Paper II (to which the reader is referred to for full details). In particular, we used the same line list (optimized for early K-type stars) and oscillator strengths. To calibrate the atomic data, we made use of a very high S/N FEROS moonlight spectrum and a Kurucz solar model with $T_{\rm eff}$=5777 K, $\log g$=4.44 cm s$^{-2}$  and a microturbulent velocity $\xi$=1.0 km s$^{-1}$ (Cox 2000). The oscillator strengths were adjusted until Kurucz's solar abundances (quoted in Table~\ref{tab_a1}, only available in electronic form) were reproduced. An extensive comparison between our $\log gf$ values and literature data is presented in the Appendix of Paper \nolinebreak II.

The effective temperature and surface gravity were derived from the excitation and ionization equilibrium of the iron lines, while the microturbulent velocity was determined by requiring the \ion{Fe}{i} abundances to be independent of the line strength. The abundances were derived from a differential LTE analysis using plane-parallel, line-blanketed Kurucz atmospheric models (with a length of the convective cell over the pressure scale height, $\alpha$=$l/H_{\rm p}$=0.5, and no overshoot) and the current version of the MOOG software (Sneden 1973). The equivalent widths (EWs) were measured from Gaussian fitting and are quoted in Table~\ref{tab_a1}. The lithium abundance was determined from a spectral synthesis of the \ion{Li}{i} $\lambda$6708 doublet (see Paper II).

\begin{figure*}
\resizebox{\hsize}{!}
{\rotatebox{0}{\includegraphics{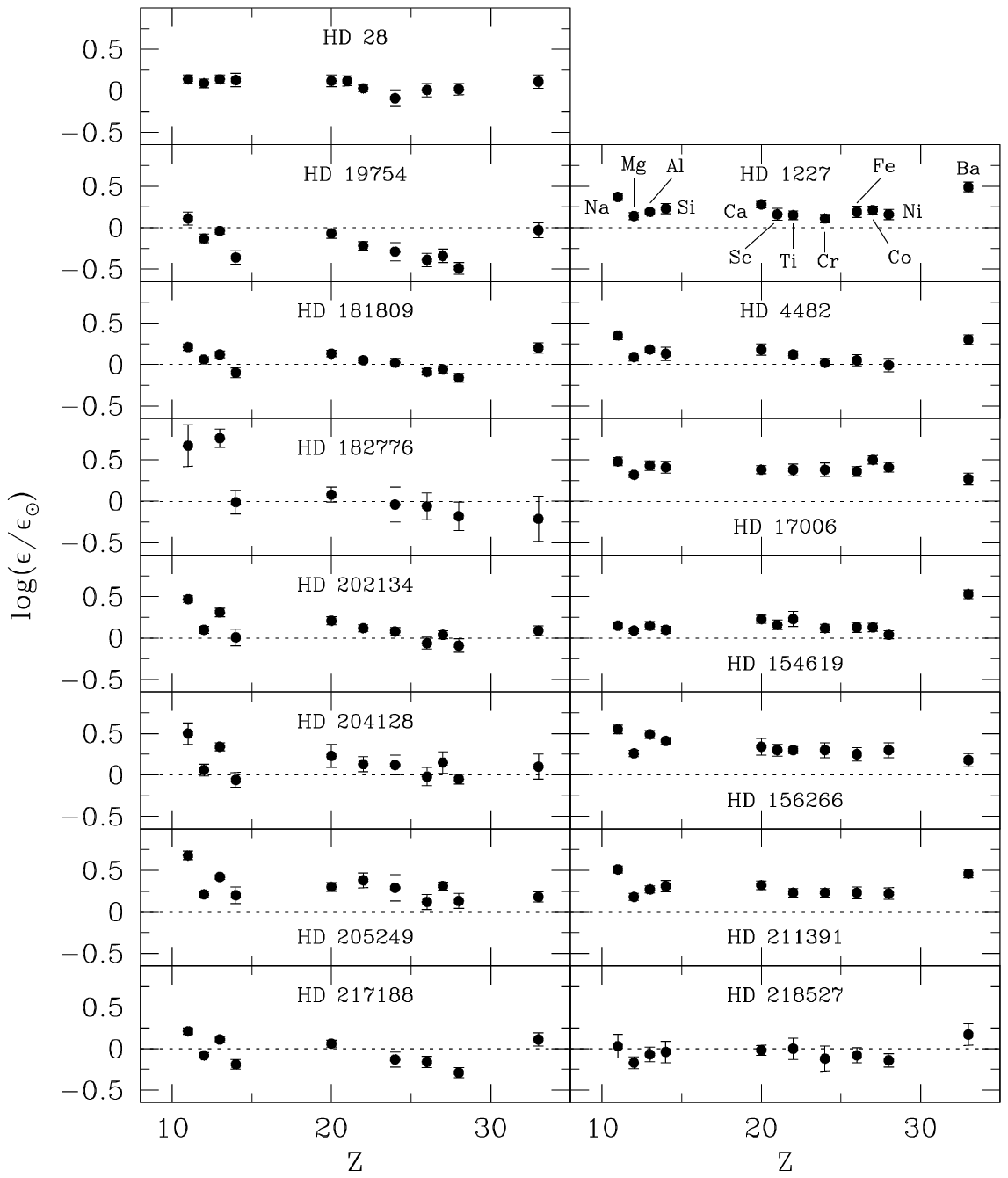}}}
\caption{Abundance patterns of the RS CVn binaries (\emph{left-hand panels}) and the stars in the control sample (\emph{right-hand panels}). The chemical elements are identified in the right-hand, upper panel. The position of barium has been shifted for the sake of clarity to $Z$=33.} 
\label{fig_pattern}
\end{figure*}

\setcounter{subtable}{1}
\begin{table*}
\caption{Abundances of the active binaries. We use the usual notation: [A/B]=$\log$ [${\cal N}$(A)/${\cal N}$(B)]$_{\star}$$-$$\log$ [${\cal N}$(A)/${\cal N}$(B)]$_{\sun}$. The corresponding 1-$\sigma$ uncertainties and the number of lines used, $N$, are indicated. Blanks indicate that no EWs for the element in question could be reliably measured. The last row gives the lithium abundances. In this case we use the usual notation: $\log
  \epsilon({\rm Li})=\log {\cal N}({\rm Li})-12.0$. The abundances of Na, Mg, Ca and Li have been corrected for departures from LTE (see text and Table~\ref{tab_b1}).}
\label{tab_2a}
\begin{center}
\begin{tabular}{lrrrrrrrr} \hline\hline
 &  \multicolumn{1}{c}{\object{HD 28}}  & \multicolumn{1}{c}{\object{HD 19754}}  & \multicolumn{1}{c}{\object{HD 181809}} &  \multicolumn{1}{c}{\object{HD 182776}} &  \multicolumn{1}{c}{\object{HD 202134}}  & \multicolumn{1}{c}{\object{HD 204128}}  & \multicolumn{1}{c}{\object{HD 205249}}  & \multicolumn{1}{c}{\object{HD 217188}}\\
\multicolumn{1}{l}{Name} & \multicolumn{1}{c}{\object{33 Psc}} & \multicolumn{1}{c}{\object{EL Eri}} & \multicolumn{1}{c}{\object{V4138 Sgr}} & \multicolumn{1}{c}{\object{V4139 Sgr}} & \multicolumn{1}{c}{\object{BN Mic}} & \multicolumn{1}{c}{\object{BH Ind}} & \multicolumn{1}{c}{\object{AS Cap}} & \multicolumn{1}{c}{\object{AZ Psc}} \\\hline
${\rm [Fe/H]}$   &   0.01$\pm$0.08 &  --0.39$\pm$0.08  &  --0.09$\pm$0.04  &  --0.06$\pm$0.16  &  --0.06$\pm$0.07  &  --0.02$\pm$0.11   &  0.12$\pm$0.09   &--0.16$\pm$0.07 \\
$N$ & 47 & 37 & 38 & 20 & 45 & 30 & 36 & 43\\
${\rm [Na/Fe]}$   &  0.13$\pm$0.05   & 0.50$\pm$0.08  &  0.30$\pm$0.04  &  0.73$\pm$0.25  &  0.53$\pm$0.04   & 0.52$\pm$0.13 &  0.56$\pm$0.05   &  0.37$\pm$0.04\\
$N$ & 1 & 1 & 1 & 1 & 1 & 1 & 1 & 1 \\
${\rm [Mg/Fe]}$    &  0.08$\pm$0.05  &  0.26$\pm$0.05  &  0.15$\pm$0.03   &   & 0.16$\pm$0.04   &  0.08$\pm$0.07  &    0.09$\pm$0.04  &   0.08$\pm$0.03\\
$N$  & 1  & 1  & 1  &   & 1  & 1  & 1  & 1  \\
${\rm [Al/Fe]}$    & 0.13$\pm$0.05  &  0.35$\pm$0.04 &   0.21$\pm$0.04  &  0.82$\pm$0.11  &  0.37$\pm$0.05  &  0.36$\pm$0.05  &  0.30$\pm$0.03   &  0.27$\pm$0.03\\
$N$ & 2 & 1 & 2 & 1 & 2 & 1 & 2 & 2\\
${\rm [Si/Fe]}$   &  0.12$\pm$0.08    & 0.03$\pm$0.08   & --0.01$\pm$0.06   &  0.05$\pm$0.14  &   0.07$\pm$0.10  &  --0.04$\pm$0.09  &    0.08$\pm$0.10  &  --0.03$\pm$0.06\\
$N$ & 9 & 5 & 6 & 4 & 6 & 4 & 7 & 5\\
${\rm [Ca/Fe]}$    & 0.11$\pm$0.07  &  0.32$\pm$0.06   & 0.22$\pm$0.04  &  0.14$\pm$0.09  &  0.27$\pm$0.05  &  0.25$\pm$0.14  &  0.18$\pm$0.05  &  0.22$\pm$0.04\\
$N$  & 3 & 3 & 3 & 3 & 3 & 3 & 2 & 3\\
${\rm [Sc/Fe]}$    & 0.11$\pm$0.06   &     &    &      &     &      &    &   \\
$N$  & 1 & & & & & & & \\
${\rm [Ti/Fe]}$     &  0.02$\pm$0.04  &  0.17$\pm$0.05   & 0.14$\pm$0.03    &   &  0.18$\pm$0.04   & 0.15$\pm$0.09    &0.26$\pm$0.09   &   \\
$N$ & 1 & 1 & 1 & & 1 & 1 & 1 & \\
${\rm [Cr/Fe]}$   &  --0.10$\pm$0.10   &  0.10$\pm$0.11   & 0.11$\pm$0.05   &  0.02$\pm$0.21  &  0.14$\pm$0.05   & 0.14$\pm$0.12   &  0.17$\pm$0.16  &   0.03$\pm$0.09\\
$N$  & 3 & 2 & 1 & 2 &1  & 2 & 3 & 3\\
${\rm [Co/Fe]}$   &       & 0.05$\pm$0.08  &   0.03$\pm$0.04   &      & 0.10$\pm$0.05  &  0.17$\pm$0.13   &  0.19$\pm$0.05  &   \\
$N$  & & 1 & 1 & & 1 & 1 & 1 & \\
${\rm [Ni/Fe]}$    &   0.01$\pm$0.07  &  --0.10$\pm$0.07   & --0.07$\pm$0.05  & --0.12$\pm$0.17  &  --0.03$\pm$0.08  &  --0.03$\pm$0.06  &   0.01$\pm$0.09  & --0.13$\pm$0.06\\
$N$  & 10 & 8 & 10 & 5 & 10 & 8 & 10 & 8\\
${\rm [Ba/Fe]}$    &  0.10$\pm$0.08  &  0.36$\pm$0.09  &  0.29$\pm$0.06  & --0.15$\pm$0.27  &  0.15$\pm$0.06  &  0.12$\pm$0.15  &   0.06$\pm$0.06  &   0.27$\pm$0.08\\
$N$  & 1 & 1 & 1 & 1 & 1 & 1 & 1 & 1\\
$\log \epsilon$(Li) & $<$0.87 & 1.20$\pm$0.13 & 1.23$\pm$0.09 & 1.18$\pm$0.22 & 0.95$\pm$0.15 & 1.49$\pm$0.17 & 2.03$\pm$0.10 & 1.22$\pm$0.12\\\hline
\end{tabular}
\end{center}
\end{table*}

\addtocounter{table}{-1}
\addtocounter{subtable}{1}
\begin{table*}
\caption{Abundances of the control stars.}
\label{tab_2b}
\begin{center}
\begin{tabular}{lrrrrrrrr} \hline\hline
 &  \multicolumn{1}{c}{\object{HD 1227}} & \multicolumn{1}{c}{\object{HD 4482}} & \multicolumn{1}{c}{\object{HD 17006}}&  \multicolumn{1}{c}{\object{HD 154619}} & \multicolumn{1}{c}{\object{HD 156266}} & \multicolumn{1}{c}{\object{HD 211391}} & \multicolumn{1}{c}{\object{HD 218527}} \\\hline
${\rm [Fe/H]}$  &   0.19$\pm$0.07  &  0.05$\pm$0.07  & 0.36$\pm$0.06 &  0.13$\pm$0.06 &  0.25$\pm$0.08 &  0.23$\pm$0.07 &   --0.08$\pm$0.09    \\
$N$  & 40 & 37 & 51 & 48 & 35 & 39 & 36\\
${\rm [Na/Fe]}$   & 0.18$\pm$0.04 &  0.30$\pm$0.05 &  0.12$\pm$0.05   & 0.02$\pm$0.04  & 0.30$\pm$0.05  & 0.28$\pm$0.04  & 0.11$\pm$0.14   \\ 
$N$  & 1 &  1 &  1 &  1 &  1 &  1 &  1 \\
${\rm [Mg/Fe]}$   & --0.05$\pm$0.04  &  0.04$\pm$0.05  & --0.04$\pm$0.03 &  --0.04$\pm$0.03  &  0.01$\pm$0.04  &  --0.05$\pm$0.04  & --0.09$\pm$0.07\\  
$N$  &  1 & 1  & 1  & 1  &  1 &  1 & 1  \\
${\rm [Al/Fe]}$ &  0.00$\pm$0.03 &  0.13$\pm$0.03  &  0.07$\pm$0.06 &   0.02$\pm$0.05  & 0.24$\pm$0.05  &  0.04$\pm$0.04  &  0.02$\pm$0.08  \\   
$N$  & 2 & 2 & 2 & 2 & 2 & 2 &1 \\
${\rm [Si/Fe]}$  &   0.04$\pm$0.06  &  0.08$\pm$0.08  &  0.05$\pm$0.07 &  --0.03$\pm$0.04 &  0.16$\pm$0.04   & 0.08$\pm$0.07  &   0.04$\pm$0.13  \\  
$N$  & 6 & 7 & 7 & 6 & 3 & 7 & 7\\
${\rm [Ca/Fe]}$   &  0.09$\pm$0.04 &  0.13$\pm$0.07   & 0.02$\pm$0.05  &  0.10$\pm$0.05 &   0.09$\pm$0.10   & 0.09$\pm$0.05   &  0.06$\pm$0.06  \\  
$N$  & 3 & 3 & 3 & 3 & 3 & 3 & 3\\
${\rm [Sc/Fe]}$ &   --0.03$\pm$0.07  &    &     & 0.03$\pm$0.06  &  0.05$\pm$0.07   &    &    \\  
$N$  & 1 & & & 1 & 1 & & \\
${\rm [Ti/Fe]}$   & --0.04$\pm$0.05  & 0.07$\pm$0.04  &  0.02$\pm$0.07   & 0.10$\pm$0.09   & 0.05$\pm$0.04 & 0.00$\pm$0.05   &  0.08$\pm$0.13  \\  
$N$  & 1 & 1 & 1 & 1 & 1 & 1 & 1\\
${\rm [Cr/Fe]}$    &--0.08$\pm$0.05 &  --0.03$\pm$0.05   & 0.02$\pm$0.08  & --0.01$\pm$0.05    &0.05$\pm$0.09 & 0.00$\pm$0.05  & --0.04$\pm$0.15  \\
$N$  & 1 & 1 & 2 & 1 & 2 & 1 & 1\\
${\rm [Co/Fe]}$  &   0.02$\pm$0.05 &     &  0.14$\pm$0.05 & 0.00$\pm$0.05 &    &     &   \\    
$N$  & 1 & & 1 & 1 & & & \\
${\rm [Ni/Fe]}$  &  --0.03$\pm$0.06 &  --0.06$\pm$0.08 &   0.05$\pm$0.06 &   --0.09$\pm$0.05 &    0.05$\pm$0.09 &   --0.01$\pm$0.07 &  --0.06$\pm$0.08  \\ 
$N$  & 10 & 10 & 10 & 10 & 10 & 10 & 8\\
${\rm [Ba/Fe]}$  &  0.30$\pm$0.06 &   0.25$\pm$0.06 &  --0.09$\pm$0.07 &   0.40$\pm$0.05 &   --0.07$\pm$0.08 &   0.23$\pm$0.05 &   0.25$\pm$0.13  \\
$N$  & 1 & 1 & 1 & 1 & 1 & 1 & 1\\
$\log \epsilon$(Li) & 1.16$\pm$0.18 & 1.31$\pm$0.12 & $<$1.18 & $<$1.03 & $<$0.75 & $<$0.95 & $<$0.51 \\\hline
\end{tabular}
\end{center}
\end{table*}

The reliability of abundance analyses relying on excitation temperatures may be questioned because of departures from LTE preferentially affecting the low-excitation iron lines (e.g. Ruland \etal 1980). Similarly to what has been done in Paper II, we have therefore repeated the whole procedure after rejection of the \ion{Fe}{i}
lines with a first excitation potential, $\chi$, below 3.5 eV. In this case, we used effective temperatures derived from $T_{\rm eff}$-($B-V$) empirical calibrations (Alonso \etal 1999). As will be shown below, these values turn out in most cases to be slightly lower than the temperatures derived from the excitation equilibrium of the \ion{Fe}{i} lines. The resulting abundance ratios will not be considered further, as they are identical, within the uncertainties, to the values derived using our main method.

\section{Results} \label{sect_results}
The atmospheric parameters ($T_{\rm eff}$, $\log g$ and $\xi$) and abundances are given in Tables 1$a$-$b$ and 2$a$-$b$, respectively. Note that the abundances of Li, Na, Mg and Ca have been converted into NLTE values. The corrections range from $\Delta \epsilon$=$\log(\epsilon)_{\rm NLTE}$--$\log(\epsilon)_{\rm LTE}$=+0.10 to +0.26 for \ion{Li}{i} $\lambda$6708 (Carlsson et al. 1994), from --0.09 to +0.06 dex for \ion{Na}{i} $\lambda$6154 and from +0.04 to +0.15 dex for \ion{Mg}{i} $\lambda$5711 (Gratton \etal 1999). For convenience, these corrections are listed on a star-to-star basis in Table~\ref{tab_b1} (we also include the stars analyzed in Paper II). For the Ca lines, we used for all the stars $\Delta \epsilon$=+0.09, +0.07 and +0.01 for \ion{Ca}{i} $\lambda$6166, 6456 and 6500, respectively (Drake 1991). For the other chemical species, no NLTE corrections for the relevant range of physical parameters are available in the literature. 

The radiative loss in the \ion{Ca}{ii} H+K lines in units of the bolometric luminosity, $R_{\rm HK}^{\prime}$, was used as primary activity indicator. This quantity was derived from our spectra following Linsky et al. (1979). The relative emission-line fluxes of the \ion{Ca}{ii} H+K lines were calculated by integrating the unnormalized spectra in the wavelength range encompassing the emission components, and subsequently dividing these quantitities by the total counts integrated in the 3925--3975 \AA \ bandpass. The conversion to absolute fluxes was performed using an empirical calibration between $(V-R)$ and the absolute surface flux in the 3925--3975 \AA \ wavelength range, ${\cal F}_{50}$ (Linsky et al. 1979). Finally, these two quantities were corrected for photospheric contribution, and the sum divided by the bolometric luminosity to obtain $R_{\rm HK}^{\prime}$ (see Paper II for further details). The colour anomalies discussed in Sect.~\ref{sect_teff} lead to a systematic, dramatic underestimation of ${\cal F}_{50}$, and ultimately $R_{\rm HK}^{\prime}$ (up to 0.5 dex), when adopting the observed $(V-R)$ colour. We therefore used instead the colour appropriate for a star with the derived $T_{\rm eff}$ and [Fe/H] (eq. (5) of Alonso \etal 1999). We regard this approach as a major improvement compared to what has been done in Paper II. Consequently, the activity indices for the 6 stars discussed in this paper have been re-calculated accordingly. The bolometric luminosities were estimated from theoretical isochrones (see Sect.\ref{sect_logg}). We also define $R_{\rm X}$, which is given as the ratio between the X-ray (in the {\em ROSAT} [0.1--2.4 keV] energy band) and bolometric luminosities. Finally, we use the maximum amplitude of the wave-like photometric variations in $V$ band, $\Delta V_{\rm wave}$, as a proxy of stellar spottedness (values from Strassmeier et al. 1993). The activity indices, as well as the source of the X-ray data, are given in Tables 1$a$-$b$.

The kinematic data for the programme stars were calculated following Paper II (Tables 1$a$-$b$). The control stars have kinematical properties typical of the thin disk population, but some RS CVn binaries have large rotational lags ($V$$\sim$--50 km s$^{-1}$) pointing to a possible thick disk membership. We follow Bensby, Feltzing \&  Lundstr\"om (2004) to estimate the probability that a star belongs to the thin/thick disk or to the halo, using the calculated velocity components with respect to the LSR ($U$, $V$ and $W$) and considering the characteristic velocity dispersions of these stellar populations ($\sigma_U$, $\sigma_V$ and $\sigma_W$). It is assumed that the galactic space velocities in the solar neighbourhood can be approximated by a Gaussian distribution. The other quantities entering in the calculations are the observed local fraction of stars, $X$, and the asymmetric drift, $V_{\rm asym}$, corresponding to each population. All the relevant values are taken from table 1 of Bensby \etal (2004). The likelihood for a thin or thick disk membership can be parameterized by the ratio of the respective probabilities for both components, $D$/$TD$ (Bensby \etal 2004). Only \object{HD 83442} is very likely to be a thick disk star ($D$/$TD$=0.02, i.e. the star is 50 times more likely to belong to the thick disk than to the thin disk), whilst the case of \object{HD 10909} and \object{HD 119285} is more ambiguous ($D$/$TD$=0.86 and 0.60, respectively). These 3 active binary systems have been discussed in Paper II. All the other stars in our sample are very likely to belong to the thin disk ($D$/$TD$ $>$ 8). None is part of the galactic halo.

\begin{figure*}
\resizebox{\hsize}{!}
{\rotatebox{0}{\includegraphics{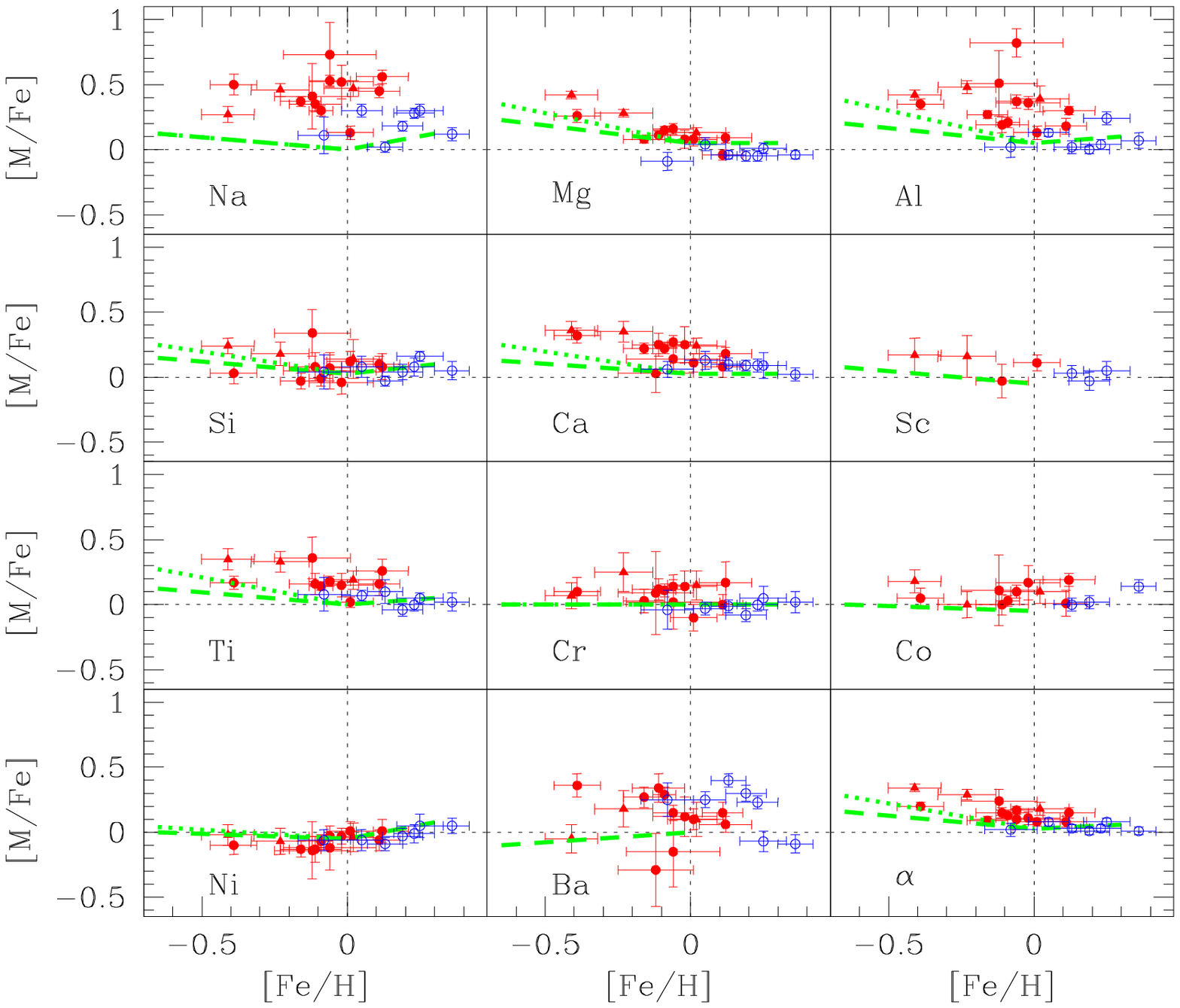}}}
\caption{Abundance ratios as a function of [Fe/H]. The active binaries and stars in the control sample are indicated by filled and open circles, respectively. The possible thick disk stars (\object{HD 10909}, \object{HD 83442} and \object{HD 119285}; see text) are indicated by filled triangles. Colour coding as in Fig.~\ref{fig_temperature}. We define the mean abundance ratio of the $\alpha$-synthezised elements, [$\alpha$/Fe], as the unweighted mean of the Mg, Si, Ca and Ti abundances. The thick dashed and dotted lines show the characteristic trends of kinematically-selected samples of thin and thick disk stars, respectively (Bensby \etal 2003), The data for Sc, Co and Ba are taken from Reddy \etal (2003). } 
\label{fig_abundance}
\end{figure*}

\section{Reliability of the atmospheric parameters} \label{sect_reliability}
An abundance analysis of active stars relying on effective temperatures and surface gravities derived from classical 1-D LTE atmospheric models might be questioned because of the existence of large-scale photospheric structures (e.g. plages, cool spots) and an overlying, prominent chromosphere. Before discussing the abundance patterns, let us first examine the reliability of our derived atmospheric parameters. In the following discussion, we shall combine our results with data taken from Paper II for 6 additional RS CVn binaries. 

\begin{figure*}
\resizebox{\hsize}{!}
{\rotatebox{0}{\includegraphics{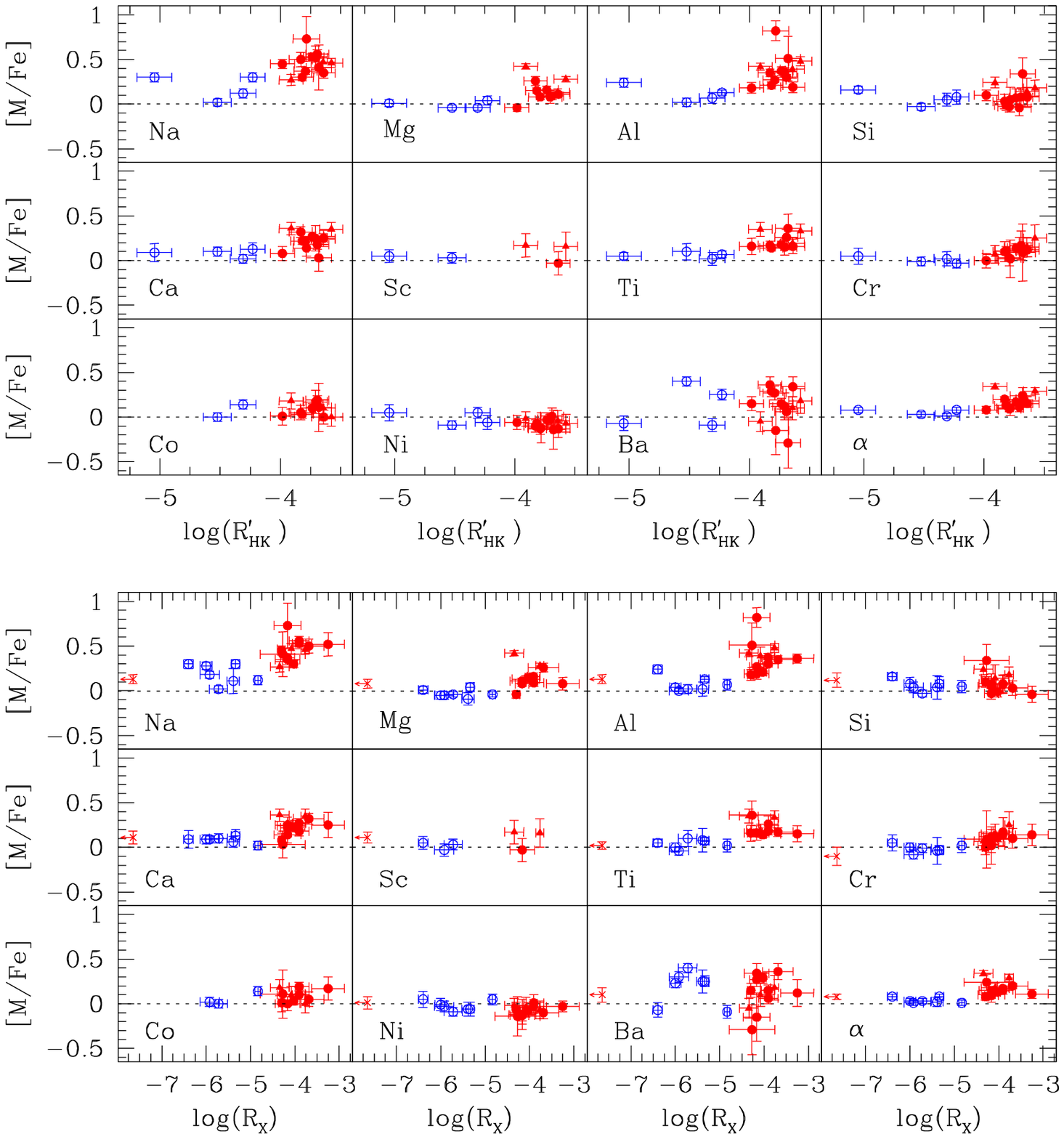}}}
\caption{Abundance ratios as a function of the activity indices $R_{\rm HK}^{\prime}$ (\emph{upper panels}) and $R_{\rm X}$ (\emph{bottom panels}). Symbols as in Fig.~\ref{fig_abundance}.} 
\label{fig_activity}
\end{figure*}

\subsection{Effective temperatures} \label{sect_teff}
A consistency check on our temperature estimates is provided by a comparison with colour temperatures computed from the ($B-V$), ($V-R$) and ($V-I$) indices, using calibrations based on the infrared flux method (Alonso \etal 1999). The photometric data used, as well as the literature sources are quoted in Tables 1$a$-$b$. Preference was given to repeated observations in order to minimize possible colour variations arising from the rotational modulation of active regions. In most cases, the adopted colours are the mean of more than 15 (and up to 70) time-resolved measurements. In case of a metallicity dependent relation, we used our derived [Fe/H] values. The excitation temperatures of the active binaries appear marginally higher than the values derived from $T_{\rm eff}$-($B-V$) calibrations: $<$$T_{\rm exc}$--$T_{\rm colour}$$>$=+80$\pm$46 K. This fair agreement is gratifying, but we cannot rule out that it is circumstantial in view of previous claims of a $(B-V)$ excess in active stars (e.g. Amado 2003; Stauffer \etal 2003). We also warn the reader that our excitation temperatures may be affected by other shortcomings (e.g. NLTE effects). In sharp contrast, the mean difference amounts to +201$\pm$36 and +215$\pm$35 K for ($V-R$) and ($V-I$), respectively. As discussed in Papers I and II, the differences are unlikely to be accounted for by spottedness or by the existence of a faint, cooler companion. 

\addtocounter{table}{0}
\addtocounter{subtable}{-2}
\begin{table*}
\centering
\caption{False alarm probabilities for a chance correlation between the temperature differences and the activity indices (all values expressed in per cent). The numbers in brackets refer to the active binary dataset alone.}
\label{tab_3}
\begin{tabular}{lccc} \hline\hline
 & $T_{\rm exc}$--$T$($B-V$) & $T_{\rm exc}$--$T$($V-R$) & $T_{\rm exc}$--$T$($V-I$) \\\hline
log ($R_{\rm HK}^{\prime}$)         & 13.7 (58.2) & 2.9 (3.3) & 5.3 (5.8) \\
log ($R_{\rm X}$) (whole dataset)$^a$   & 7.4 (95.1) & 0.2 (4.6) & 0.1 (1.2) \\
log ($R_{\rm X}$) (detections only) & 12.7 (95.1) & 0.3 (4.6) & 0.3 (1.2) \\
$\Delta V_{\rm wave}$               & (41.0) & (19.9) & (8.9)\\\hline
\end{tabular}\\
$^a$ We used the Kendall's $\tau$ method in the case of censored data (\object{HD 28} is undetected in X-rays).
\end{table*}

We show in Fig.~\ref{fig_temperature} a comparison between the excitation and colour temperatures, as a function of the three activity indices defined above. There is a general trend for an increasing disagreement with higher $R_{\rm HK}^{\prime}$ and $R_{\rm X}$ values. The false alarm probabilities for a chance correlation between the temperature differences and the activity indices are quoted in Table~\ref{tab_3}. An increasing discrepancy between the excitation temperatures and the ($V-R$)- and ($V-I$)-based estimates at higher activity level is quantitatively confirmed for the $R_{\rm X}$ and, to a lesser extent, for the $R_{\rm HK}^{\prime}$ data. This conclusion holds irrespective of whether the whole dataset or only the active binaries are considered. There is a hint of a similar trend for the ($B-V$) data, but the correlation is not statistically significant. 

The tendency for the ($V-R$) and ($V-I$) indices to yield systematically lower temperatures than ($B-V$) has been reported in several occasions for active stars, either in the field (Fekel, Moffett \& Henry 1986) or in young open clusters (Garc\'{\i}a L{\'o}pez, Rebolo \& Mart\'{\i}n 1994; Randich \etal 1997; Soderblom \etal 1999). Although early attempts to link these colour anomalies to activity have been inconclusive in disk stars, presumably because of a limited range in activity level (Ruci\'nsky 1987), evidence for a ($V-I$) excess in the most active Pleiades stars has been presented by King, Krishnamurthi \& Pinsonneault (2000). A distortion of the spectral energy distribution because of activity-related processes is indeed expected, but the quantitative effect on the broad band colours is difficult to assess (e.g. Stuik, Bruls \& Rutten 1997). Chromospheric models for M-type dwarfs indicate a substantial modulation, as a function of the activity level, not only of the emergent continuum flux in the UV domain, but also in the optical and near-infrared wavebands (Houdebine \etal 1996; see, however, West \etal 2004). On the observational side, the infrared excess intrinsic to some RS CVn binaries has been claimed to diagnose the existence of cool, circumstellar material (e.g. Scaltriti \etal 1993, and references therein).

\subsection{Surface gravities} \label{sect_logg}
We show in Fig.~\ref{fig_iso} the position of our stars in HR diagrammes, along with theoretical isochrones chosen to match as closely as
possible our derived [Fe/H] and [$\alpha$/Fe] values (Yi et al. 2001; Kim et al. 2002). We use {\em Hipparcos} data and the excitation temperatures. The stellar ages and masses derived from the evolutionary tracks are given in Tables 1$a$-$b$. The surface gravities determined from the ionization
equilibrium of the Fe lines appear systematically lower for the RS CVn binaries than the values derived from the isochrones ($<$$\log g_{\rm ioni}$--$\log g_{\rm iso}$$>$=--0.21$\pm$0.06 dex). In contrast, there is a good agreement for the control stars: $<$$\log g_{\rm ioni}$--$\log g_{\rm iso}$$>$=--0.01$\pm$0.08 dex. This discrepancy in the gravity estimates could be plausibly caused by an overionization of iron (see below). There is a hint that the surface gravities determined from fitting the wings of collisionally-broadened lines tend to be systematically higher than values derived from the ionization equilibrium of the Fe lines (Paper I). Other interpretations are, however, possible. Because $T_{\rm eff}$ and $\log g$ are interdependent, for instance, such a gravity offset could be an artefact of a small, systematic underestimation of the effective temperatures (at the $\sim$70 K level).

Summarizing, we conclude that our $T_{\rm eff}$ and $\log g$ estimates are liable to systematic errors of the order of 150 K and 0.25 dex, respectively. The sensitivity of the abundance ratios against such changes is examined in Table~\ref{tab_4}. The impact is limited and is typically comparable to the abundance uncertainties ($\Delta[{\rm M/Fe}]\la0.12$ dex).

\section{Abundance patterns} \label{sect_abundances}
As can be seen in Fig.~\ref{fig_pattern}, there is a fairly clear dichotomy between the abundance patterns of the two samples of active and inactive stars. The active binaries exhibit as a class a chemical composition departing more conspicuously from the solar mix (compare the left- and right-hand panels). 

Figure~\ref{fig_abundance} shows the abundance ratios as a function of [Fe/H]. The increasing abundance ratios of several chemical species (e.g. the $\alpha$ elements) with decreasing [Fe/H] is a well-known feature of galactic disk stars, and may be naturally explained in terms of a varying relative contribution of Type Ia/Type II supernova ejecta to the chemical composition of the natal interstellar cloud. However, the large overabundances observed cast some doubt on this interpretation. We overplot in Fig.~\ref{fig_abundance} the characteristic abundance patterns of kinematically-selected samples of thin and thick disk FG dwarfs (Bensby \etal 2003). The overabundances presented by the active binaries appear generally much more marked at a given metallicity than for either of these two populations. For instance, [Ca/Fe]$\sim$0.35 dex at [Fe/H]=--0.4, while [Ca/Fe]$\sim$0.1 and 0.2 dex only for thin and thick disk stars, respectively (we recall that very few stars in our sample are expected to belong to the thick disk; Sect.~\ref{sect_results}). The dramatic Al overabundances at near solar metallicities (up to 0.8 dex) are also completely at variance with what is expected for stars in the solar neighbourhood. The origin of the high Na and Ba abundances in some control stars is unclear, but might be related to systematic errors in the atomic data. The abundances are derived from a single line in both cases. 

The RS CVn systems often display higher abundance ratios than the control stars at a given [Fe/H] value (Fig.~\ref{fig_abundance}). To test the hypothesis that it is related to differences in the activity level, we plot in Fig.~\ref{fig_activity} the abundance ratios as a function of $R_{\rm HK}^{\prime}$ and $R_{\rm X}$. The positive correlation seen for several elements (e.g. Al, Ca) suggests that the observed overabundances may arise from activity-related processes. This hypothesis is examined in the following. 

\section{Discussion}\label{sect_discussion}
\begin{figure}
\resizebox{\hsize}{!}
{\rotatebox{0}{\includegraphics{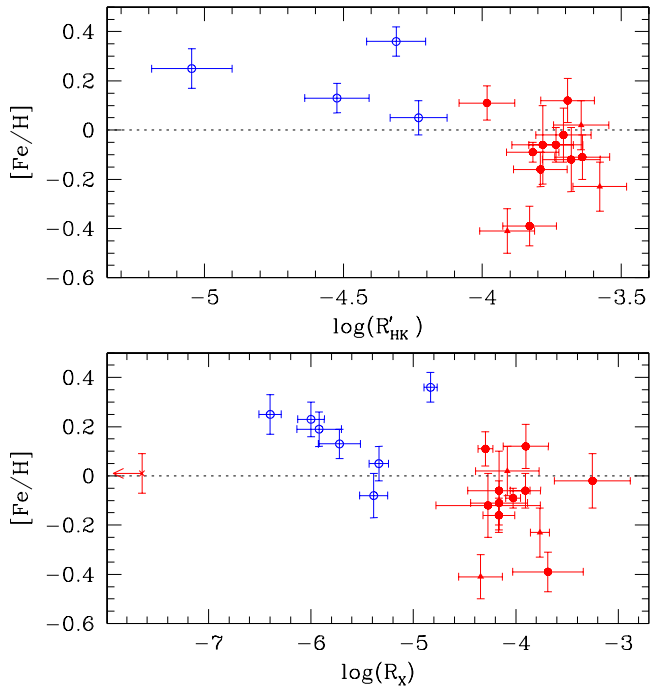}}}
\caption{Iron content as a function of the activity indices $R_{\rm HK}^{\prime}$ (\emph{upper panel}) and $R_{\rm X}$ (\emph{bottom panel}). Symbols as in Fig.~\ref{fig_abundance}.} 
\label{fig_fe}
\end{figure}

Cool spot groups are known to cover a large fraction of the stellar photosphere in chromospherically active stars. For one star in our sample (\object{HD 119285}; see Paper II), Saar, Nordstr\"om \& Andersen (1990) derived a spot filling factor in the range 0.18-0.37 from a modelling of the TiO bandhead at 8860 \AA. To investigate the impact of these features on our results, we have created a composite Kurucz synthetic spectrum of three fictitious stars with $T_{\rm eff}$=4830 K, covered by cooler regions ($\Delta T_{\rm eff}$=--1000 K) with different covering factors, $f_s$=0, 30 and 50\% (Paper II, see also Neff, O'Neal \& Saar 1995). We then derived the abundances from these composite spectra following exactly the procedure used for the analysis of the active binaries. The differences between the results obtained for the unspotted star are given in Table~\ref{tab_5}. As can be seen, the existence of cool spots generally leads to a systematic overestimation of the abundance ratios with respect to iron. The impact on the resulting abundances can be significant and may account for the overabundances observed for some elements (e.g. Ca). This mostly results from a filling in of the iron line profiles, a behaviour which seems supported by the apparent iron deficiency at high activity level seen in Fig.~\ref{fig_fe} (see also Cayrel, Cayrel de Strobel \& Campbell 1985; Drake \& Smith 1993).\footnote{We note, however, that the two samples of active and inactive stars are likely drawn from two distinct populations in terms of stellar ages. The median evolutionary age of the control sample is only about 0.6 Gyr, against 2.4 Gyr for the active binaries. Based on the statistical relationship in galactic disk stars between the stellar age and the iron content (Rocha-Pinto \etal 2000), we might expect a $\sim$0.2 dex offset in the mean metallicity of the two samples.} If this interpretation is correct, an asymmetrical distribution of cooler photospheric regions is expected to lead to a variation of the derived abundances as a function of the rotational phase. Ottmann, Pfeiffer \& Gehren (1998) carried out time-resolved spectroscopy of several RS CVn binaries, but did not find convincing evidence for such an effect. 

Most spectral lines used in this study are generally strong and of low excitation, and are therefore bound to form in upper photospheric layers where the effect of the low chromosphere may be appreciable. We have investigated in Paper II the influence of a chromospheric temperature rise on the abundance ratios. To this end, a LTE abundance analysis was carried out using modified Kurucz atmospheric models characterized by a temperature gradient reversal in the outermost regions and an overall heating (up to 190 K) of the deeper photospheric layers (see fig.3 of Paper II). We considered two model chromospheres meant to be representative of K-type giants (Kelch et al. 1978) and RS CVn binaries (Lanzafame, Bus\`a \& Rodon\`o 2000). Excluding Sc and Ba, the incorporation of a chromospheric component yields abundance ratios that do not differ by more than 0.13 dex from those given by classical Kurucz models. Taking into account chromospheric heating may have a significant impact on the derived abundances of some elements, but appears insufficient to account for the anomalously high Na and Al abundances, for instance. 

Another important, related issue is the fact that the NLTE corrections applied to the abundances of some elements (Na, Mg, Ca) are based on 'conventional' atmospheric models not including a chromospheric component, the latter being a key ingredient in NLTE line formations studies (e.g. Bruls, Rutten \& Shchukina 1992). In a companion paper, we have recently discussed the behaviour of the optical oxygen abundance indicators in active stars (Morel \& Micela 2004). While [\ion{O}{i}] $\lambda$6300 appears to be largely unaffected by chromospheric activity, the 7774 \AA-\ion{O}{i} triplet yields abundances dramatically increasing with $R_{\rm HK}^{\prime}$ or $R_{\rm X}$, even after 'canonical' NLTE corrections are applied. These two diagnostics have about the same sensitivity (in an absolute sense) against atmospheric parameter changes, but a completely distinct behaviour in terms of departures from LTE (the \ion{O}{i} triplet is strongly affected, whereas [\ion{O}{i}] $\lambda$6300 is insensitive to these effects). This discrepancy between the two indicators may therefore indicate that the NLTE corrections to apply to the permitted oxygen lines in chromospherically active stars largely exceed what is anticipated for their inactive analogues. It is conceivable that a similar conclusion, albeit to a lesser extent, also holds for other chemical species. As far as iron is concerned, however, the fair agreement between the excitation/($B$--$V$)-based temperatures or the ionization/isochrone gravities (Sect.~\ref{sect_reliability}) does not argue in favour of strong deviations from LTE in our sample. One may expect overionization to be an important process in active stars in view of their strong UV excess (Stauffer \etal 2003), and to manifest itself by an overall depletion of the neutral ionization stage. Evidence for an unexpectedly large overionization of iron in young open cluster stars is indeed recently emerging (e.g. Schuler \etal 2003; Yong \etal 2004).  If explained in terms of NLTE effects, the large overabundances derived from neutral lines (e.g. \ion{Na}{i}, \ion{Al}{i}) would apparently require a competing physical mechanism capable of dramatically overpopulating the neutral stage in such a way that it overwhelms the action of overionization (see, e.g. Bruls \etal 1992 for a discussion of the various NLTE mechanisms possibly at work).

\begin{table}
\caption{Effect of changes in the stellar parameters on the abundance ratios derived for \object{HD 10909} (see Paper II). The iron content is given by the \ion{Fe}{i} lines.}
\begin{center}
\label{tab_4}
\begin{tabular}{lccc} \hline\hline
                & $\Delta T_{\rm eff}$=+150 & $\Delta \log g$=+0.25 dex & $\Delta \xi$=+0.20 \\
                & (K) & (cm s$^{-2}$) & (km s$^{-1}$) \\\hline
$\Delta$${\rm [Fe/H]}$        & +0.03  & +0.04  & --0.08\\
$\Delta$${\rm [Na/Fe]}$       & +0.06  & --0.05 & +0.05 \\
$\Delta$${\rm [Mg/Fe]}$       & +0.01  & --0.06 & --0.02\\
$\Delta$${\rm [Al/Fe]}$       & +0.04  & --0.04 & +0.06 \\
$\Delta$${\rm [Si/Fe]}$       & --0.12 & +0.01  & +0.04 \\
$\Delta$${\rm [Ca/Fe]}$       & +0.05  & --0.05 & --0.04\\
$\Delta$${\rm [Sc/Fe]}$       & --0.06 & +0.07  & +0.07 \\
$\Delta$${\rm [Ti/Fe]}$       & +0.08  & --0.04 & +0.06 \\
$\Delta$${\rm [Cr/Fe]}$       & +0.06  & --0.03 & +0.03 \\
$\Delta$${\rm [Co/Fe]}$       & +0.01  & +0.01  & +0.05 \\
$\Delta$${\rm [Ni/Fe]}$       & --0.04 & +0.02  & +0.02 \\
$\Delta$${\rm [Ba/Fe]}$       & --0.09 & +0.09  & --0.10\\
$\Delta$${\rm [\alpha/Fe]}$   & +0.00  & --0.04 & +0.01 \\\hline
\end{tabular}
\end{center}
\end{table}

Contrary to the situation for some globular cluster stars ascending the red giant branch, it is unlikely that the surface abundance anomalies observed result from deep mixing processes not accounted for by standard evolutionary models. One could speculate that this extra mixing mechanism is related in some way to the spinning up of the stars in asynchronous X-ray binaries because of the transfer of orbital angular momentum (see, e.g. Denissenkov \& Tout 2000). This hypothesis cannot be discarded on the basis of the generally high Li abundances derived (Table~\ref{tab_2a}; Paper II), as little Li depletion might not provide compelling evidence against prior mixing phases (Charbonnel \& Balachandran 2000).\footnote{No relationship was found in our sample between the lithium content and various quantities, including the evolutionary age, the activity level or the rotational velocity (see Paper II in the latter case).} This extra mixing leads in some red giants to an enhancement of the O and Mg content accompanied by a decrease of the Al and Na abundances (e.g. Kraft \etal 1997), but these observational signatures are not seen in our sample. 

Independently from the various activity processes discussed above, we call attention to the fact that temperature effects may also play a significant role in inducing abundance peculiarities. In close analogy to the behaviour of the oxygen triplet (Morel \& Micela 2004), the abundance ratios of several elements are anticorrelated with $T_{\rm eff}$ (Fig.~\ref{fig_T}). Similar trends for some chemical species, but over a wider temperature range (4600--6400 K), have been reported in a sample of planet-host stars by Bodaghee \etal (2003). The dependence with respect to $T_{\rm eff}$ is qualitatively different, however, being even opposite in some cases to what we observe (e.g. Ca). Since the coolest stars in our sample (i.e. the RS CVn binaries) are also the most active, it is very difficult to disentangle temperature and activity effects. Unfortunately, few inactive stars with near-solar metallicity and in the relevant temperature range ($T_{\rm eff}$ $\la$ 5000 K) have abundances reported in the literature (e.g. Feltzing \& Gustafsson 1998; Affer \etal 2004; Allende Prieto \etal 2004 for the most recent works). A systematic, homogeneous abundance study of a large sample of inactive K-type stars seems worthwhile for a proper interpretation of Fig.~\ref{fig_T}. Of relevance is the substantial overionization of some metals recently reported in K-type field stars (e.g. Allende Prieto \etal 2004). This deviation from ionization equilibrium may indicate caveats in our modelling of the atmospheres of cool stars or, perhaps more likely, a still fragmentary understanding of line formation in these objects. In particular, the increasing magnitude of overionization effects with decreasing temperature is not predicted by current NLTE calculations (e.g. Th\'evenin \& Idiart 1999). 

\section{Conclusion} \label{sect_conclusions}
The analysis of this enlarged sample of RS CVn binaries has allowed us to place the conclusions presented in Paper II on a much firmer footing. In particular, we confirm the existence in chromospherically active stars of a ($V-R$) and ($V-I$) excess leading to spuriously low colour temperatures. 

\begin{table}
\caption{Effect of cool spots on the derived stellar parameters and abundances. The quoted differences are given with respect to the results obtained for a Kurucz synthetic spectrum calculated for the physical parameters and iron content of \object{HD 10909} (see Paper II).}
\begin{center}
\label{tab_5}
\begin{tabular}{lccc} \hline\hline
                & $f_s$=30\% & $f_s$=50\%\\\hline
$\Delta$$T_{\rm exc}$ (K) & --100  & --200\\
$\Delta$$\log g$ (cm s$^{-2}$)    & --0.21 & --0.40\\
$\Delta$$\xi$ (km s$^{-1}$)   & --0.00 & --0.01\\
$\Delta$$\rm [Fe/H]$        & --0.09 & --0.18\\
$\Delta$$\rm [Na/Fe]$       & +0.09  & +0.19 \\
$\Delta$$\rm [Mg/Fe]$       & +0.03  & +0.04 \\
$\Delta$$\rm [Al/Fe]$       & +0.08  & +0.16 \\
$\Delta$$\rm [Si/Fe]$       & +0.00  & +0.00 \\
$\Delta$$\rm [Ca/Fe]$       & +0.12  & +0.24 \\
$\Delta$$\rm [Sc/Fe]$       & +0.01  & +0.03 \\
$\Delta$$\rm [Ti/Fe]$       & +0.07  & +0.14 \\
$\Delta$$\rm [Cr/Fe]$       & +0.06  & +0.11 \\
$\Delta$$\rm [Co/Fe]$       & +0.01  & +0.01 \\
$\Delta$$\rm [Ni/Fe]$       & +0.00  & +0.00 \\
$\Delta$$\rm [Ba/Fe]$       & +0.01  & +0.03 \\
$\Delta$$\rm [\alpha/Fe]$   & +0.06  & +0.11 \\\hline
\end{tabular}
\end{center}
\end{table}

The active binary systems analyzed are characterized by abundance ratios generally at odds with the global patterns exhibited by galactic disk stars. We have argued that systematic errors in the determination of the atmospheric parameters are not expected to have a very significant impact on the abundance patterns (Table~\ref{tab_4}). The effect of cool spot groups is to lead to a systematic overestimation of the abundance ratios with respect to iron (Table~\ref{tab_5}), in qualitative agreement with the general pattern observed. As such, this constitutes an attractive explanation for the abundance peculiarities discussed in this paper. The observed correlation between the abundance ratios of some elements and the stellar activity level (Fig.~\ref{fig_activity}) may also fit into this picture, considering the statistical relationship between the incidence of starspots and the X-ray emission (Messina \etal 2003). However, the lack of an a priori knowledge of, e.g. the temperature difference between these cool features and the undisturbed photosphere or the spot covering factor makes difficult a quantitative foray into the relevance of this scenario. In addition, the effect of other competing mechanisms (e.g. plage areas, chromospheric heating) should, in principle, also be simultaneously taken into account. The anomalous abundances presented by some elements (e.g. Na, Al) may betray the action of an additional process causally linked to chromospheric activity. Unexpectedly large departure from LTE arising from nonthermal line excitation is a plausible candidate, but the physical process capable of inducing such an overpopulation of the neutral stage remains to be identified. Overionization would be instead  expected to play a significant role in stars with such a UV excess.

\begin{figure*}
\resizebox{\hsize}{!}
{\rotatebox{0}{\includegraphics{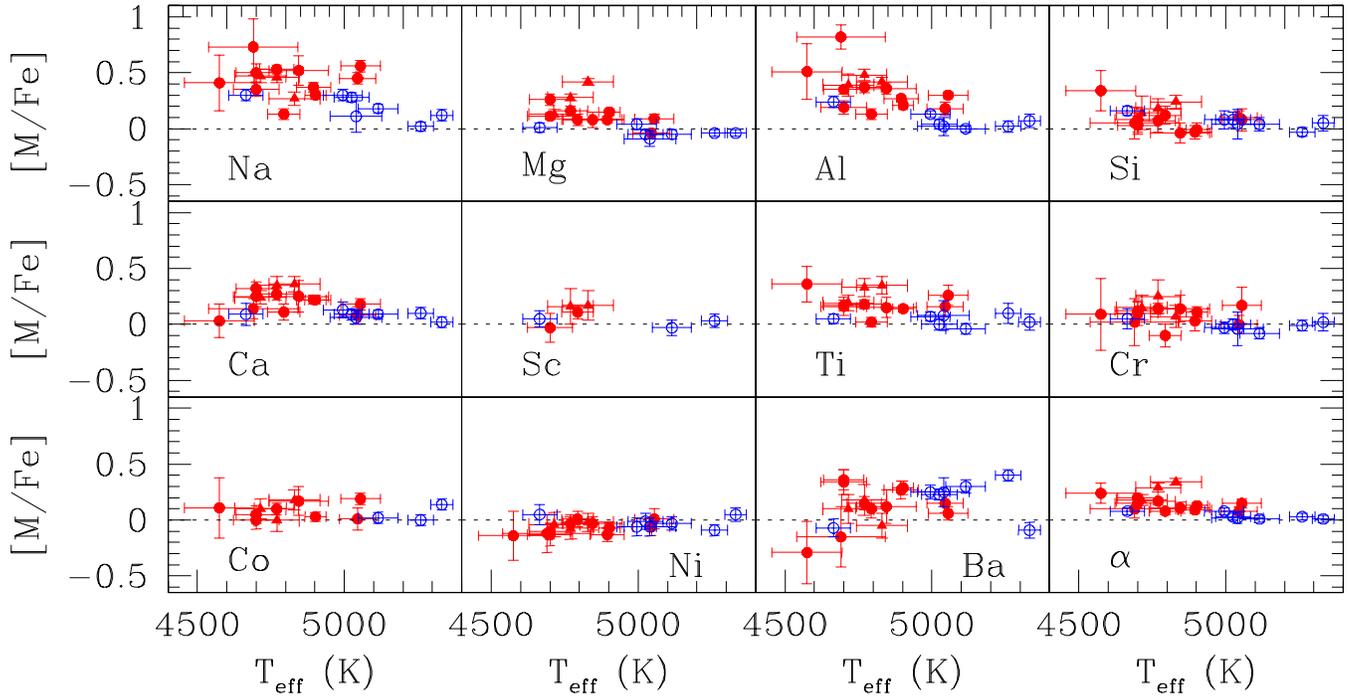}}}
\caption{Abundance ratios as a function of the effective temperatures. Symbols as in Fig.~\ref{fig_abundance}.} 
\label{fig_T}
\end{figure*}

Finally, it cannot be ruled out that {\em general} limitations in our current understanding of K-type stars with extended atmospheres (NLTE line formation, atmospheric modelling; Fields \etal 2003) are at least partly responsible of the odd abundances observed (Fig.~\ref{fig_T}). 

Determinations of the coronal abundances in chromospherically active stars have flourished in recent years (e.g. Sanz-Forcada, Favata \& Micela 2004) and have given an impetus to studies of chemical fractionisation processes in stellar coronae (e.g. Drake 2003). Unfortunately, the lack of photospheric abundance determinations for elements other than iron (which results in the customary use of the solar pattern as reference) may lead to misleading results and impedes a clear interpretation of this phenomenon. Although our initial goal when this project was initiated was to fill this caveat by determining accurate photospheric abundances for chemical species of prime interest in the context of X-ray studies (namely, O, Al, Ca, Ni, Mg, Fe, Si),\footnote{Carbon and nitrogen spectral lines are not measurable in our spectra owing to their intrinsic weakness and severe blending.} we have shown here that the photospheric abundances derived in very active stars are likely plagued by systematic effects (see Morel \& Micela 2004 in the case of oxygen). Until these effects are better understood, great caution must therefore be exercised when using these data.

It is remarkable that all our programme stars (see also Paper II) share about the same evolutionary status, i.e. they are at the bottom of the red giant branch or are starting to ascend it (Fig.~\ref{fig_iso}). This supports the idea that enhanced chromospheric activity is associated with a specific evolutionary stage during which the spinning up of the stars, coupled with the formation of a deep convective envelope, greatly enhances dynamo action (Barrado \etal 1994). We should point out, however, that our sample is not free from selection effects: we have preferentially selected slow rotators, which are presumably characterized by a comparatively low level of coronal activity.

\begin{acknowledgements}
  This research was supported through a European Community Marie Curie
  Fellowship (No. HPMD-CT-2000-00013). G.\,M. acknowledges financial support from MIUR (Ministero della Istruzione,
  dell'Universit\`a e della Ricerca). We wish to thank the anonymous referee for a careful reading of the manuscript and very useful comments, as well as Matteo Guainazzi for his kind assistance at the telescope. This research made use of NASA's Astrophysics Data System Bibliographic Services and the Simbad database operated at CDS, Strasbourg, France.  
\end{acknowledgements}

\appendix
\section{EW measurements} \label{sect_a1}
Table~\ref{tab_a1} presents the line list, oscillator strengths and EW measurements (only available  in electronic form).

\begin{table*}
\caption{Line list, calibrated atomic data and EW measurements. The adopted solar abundances are quoted for each element. A blank indicates that the EW was not reliably measurable, either because the line was affected by cosmic rays/telluric features or significantly van der Walls broadened (see Paper II), or because the Gaussian fit was judged unsatisfactory.}
\label{tab_a1}
\hspace*{-1.3cm}
\scriptsize
\begin{tabular}{lcrrrrrrrrrrrrrrrr} \hline\hline
  $\lambda$ (\AA) & $\chi$ (eV)  & log $gf$ & \multicolumn{15}{c}{EW (m\AA)}\\
          &        &           &\object{HD 28}    & \object{HD 1227} &\object{HD 4482}  &\object{HD 17006} &\object{HD 19754} &\object{HD 154619}&\object{HD 156266}&\object{HD 181809}&\object{HD 182776}&\object{HD 202134}&\object{HD 204128}&\object{HD 205249}&\object{HD 211391}&\object{HD 217188}&\object{HD 218527}\\\hline
{\bf \ion{Na}{i}}; $\log$ $\epsilon_{\odot}$=6.33&&&&&&&&&&&&&&&&\\
6154.226  & 2.102  & --1.637   &  78.2   &  80.9   &  87.7   &  86.4   &  82.9   &  68.3   & 121.3   &  82.8   & 158.0   & 111.8   & 114.0   & 120.2   &  98.8   &  87.2   &  59.1 \\
{\bf \ion{Mg}{i}}; $\log$ $\epsilon_{\odot}$=7.49&&&&&&&&&&&&&&&&\\
5711.088  & 4.346  & --1.514   & 138.6   & 134.2   & 138.5   & 141.8   & 137.3   & 124.3   & 160.1   & 144.1   &         & 153.0   & 150.7   & 156.4   & 142.7   & 140.7   & 114.3 \\
{\bf \ion{Al}{i}}; $\log$ $\epsilon_{\odot}$=6.47&&&&&&&&&&&&&&&&\\
6698.673  & 3.143  & --1.843   &  60.2   &  51.1   &  55.9   &  57.6   &  55.2   &  44.0   &  89.8   &  55.3   &         &  73.8   &  76.4   &  73.8   &  59.2   &  58.2   &  38.1 \\
7835.309  & 4.022  & --0.663   &  71.7   &  69.7   &  72.6   &  89.3   &         &  64.1   & 107.3   &  75.3   & 151.5   &  94.5   &         &  97.0   &  80.3   &  74.1   &       \\
{\bf \ion{Si}{i}}; $\log$ $\epsilon_{\odot}$=7.55&&&&&&&&&&&&&&&&\\
5793.073  & 4.930  & --1.894   &  64.1   &  68.3   &  67.9   &  72.5   &         &  60.7   &  82.5   &  56.1   &  86.0   &  66.7   &  63.9   &  75.7   &  79.5   &  59.3   &  55.8 \\
5948.541  & 5.083  & --1.098   &  95.1   & 106.6   & 102.5   & 111.6   &  92.4   &         &         &  92.8   & 118.4   &         &         &         & 113.4   &  92.6   &  93.5 \\
6029.869  & 5.984  & --1.553   &  28.8   &         &         &         &         &         &         &         &         &         &         &         &  45.9   &         &  28.0 \\
6155.134  & 5.620  & --0.742   &  87.7   & 103.7   &  99.9   & 116.9   &  76.9   &  92.1   &         &  85.2   & 101.5   &  89.0   &  82.1   & 102.0   & 111.3   &  78.0   &  78.3 \\
6721.848  & 5.863  & --1.100   &  57.2   &         &  67.4   &  76.9   &         &  57.5   &         &         &         &  60.1   &         &  74.5   &         &         &  56.5 \\
7034.901  & 5.871  & --0.779   &  65.4   &  82.8   &  77.4   &  89.5   &  51.5   &  74.2   &         &  62.1   &         &  67.3   &         &  85.9   &  87.1   &  60.2   &  62.6 \\
7680.266  & 5.863  & --0.609   &  77.5   &  97.0   &  90.9   & 109.8   &  64.5   &  86.8   & 101.4   &  77.1   &         &  80.0   &  80.2   &  94.9   &  98.2   &  76.8   &  89.0 \\
7760.628  & 6.206  & --1.356   &  28.7   &         &         &         &         &         &         &         &         &         &         &  36.8   &         &         &       \\
8742.446  & 5.871  & --0.448   &  92.9   & 112.9   & 112.4   & 118.6   &  75.4   & 105.6   & 115.4   &  82.5   & 103.8   &  88.0   &  90.7   & 114.3   & 123.3   &         &       \\
{\bf \ion{Ca}{i}}; $\log$ $\epsilon_{\odot}$=6.36&&&&&&&&&&&&&&&&\\
6166.439  & 2.521  & --1.074   & 109.6   & 107.5   & 108.7   & 105.2   & 119.3   &  99.7   & 134.7   & 116.3   & 157.0   & 129.0   & 125.1   & 128.5   & 117.5   & 118.6   &  90.1 \\
6455.598  & 2.523  & --1.350   & 103.4   &  98.4   & 103.9   &  96.6   & 109.8   &  89.9   & 133.4   & 104.5   & 148.9   & 121.5   & 128.8   &         & 109.8   & 107.1   &  83.0 \\
6499.650  & 2.523  & --0.839   & 128.2   & 128.9   & 132.5   & 127.4   & 139.1   & 116.4   & 164.9   & 142.6   & 184.6   & 155.6   & 155.8   & 155.7   & 138.1   & 143.8   & 109.9 \\
{\bf \ion{Sc}{ii}}; $\log$ $\epsilon_{\odot}$=3.10&&&&&&&&&&&&&&&&\\
6320.851  & 1.500  & --1.747   &  35.3   &  33.4   &         &         &         &  29.2   &  50.7   &         &         &         &         &         &         &         &       \\
{\bf \ion{Ti}{i}}; $\log$ $\epsilon_{\odot}$=4.99&&&&&&&&&&&&&&&&\\
5766.330  & 3.294  &   0.370   &  37.5   &  31.6   &  35.0   &  35.1   &  31.2   &  30.3   &  61.3   &  36.3   &         &  46.7   &  44.2   &  50.6   &  39.5   &         &  26.9 \\
{\bf \ion{Cr}{i}}; $\log$ $\epsilon_{\odot}$=5.67&&&&&&&&&&&&&&&&\\
5787.965  & 3.323  & --0.138   &  81.7   &  77.9   &  81.2   &  81.6   &  84.5   &  72.7   & 108.0   &  88.3   &         &  98.9   &  97.4   &  94.3   &  90.4   &  87.1   &  66.7 \\
6882.475  & 3.438  & --0.238   &  69.3   &         &         &         &         &         &         &         & 109.9   &         &         & 103.1   &         &  76.1   &       \\
6882.996  & 3.438  & --0.305   &  63.3   &         &         &         &         &         &         &         &  97.4   &         &         &  92.9   &         &  64.5   &       \\
6925.202  & 3.450  & --0.227   &         &         &         &  78.2   &  72.0   &         & 107.3   &         &         &         &  94.7   &         &         &         &       \\
{\bf \ion{Fe}{i}}; $\log$ $\epsilon_{\odot}$=7.67&&&&&&&&&&&&&&&&\\
5543.937  & 4.218  & --1.155   &  88.1   &  92.6   &  93.9   &  92.2   &  91.0   &  86.7   & 110.2   &  94.3   & 122.6   & 100.8   & 101.2   & 102.0   &  98.5   &  94.9   &  80.1 \\ 
5638.262  & 4.221  & --0.882   & 106.5   & 111.1   & 115.5   & 113.5   & 108.7   & 100.7   & 134.0   & 109.6   & 155.4   & 119.5   & 112.0   & 122.8   & 122.1   & 116.6   &  97.9 \\
5679.025  & 4.652  & --0.863   &  79.7   &  85.5   &  83.2   &         &  78.2   &         &         &  85.0   &         &  88.9   &  87.5   &  96.6   &  90.4   &  85.1   &  71.0 \\
5732.275  & 4.992  & --1.473   &  37.1   &  43.7   &         &  42.7   &         &  39.5   &  52.7   &  34.2   &  40.3   &  38.8   &  43.6   &         &  48.9   &  33.9   &       \\
5806.717  & 4.608  & --0.984   &  83.1   &  88.2   &  88.0   &  91.1   &  76.4   &  77.3   & 105.7   &  83.7   &         &  91.1   &  90.6   &  97.5   &  97.8   &  86.2   &  72.1 \\
5848.123  & 4.608  & --1.282   &         &  77.7   &         &  77.4   &  65.8   &  68.3   &         &         & 103.6   &  78.0   &  80.3   &  86.2   &         &  73.9   &  63.4 \\
5855.091  & 4.608  & --1.681   &  44.9   &  49.3   &  48.9   &  48.3   &         &  42.2   &  65.5   &  41.8   &  64.0   &  47.5   &         &  50.9   &  54.3   &  45.3   &  40.5 \\
5905.689  & 4.652  & --0.860   &  81.2   &         &         &  92.3   &         &         & 100.2   &         &         &  88.8   &  88.0   & 104.8   &         &         &       \\
5909.970  & 3.211  & --2.731   &         &         &         &  75.9   &  82.4   &  69.0   &         &  76.4   &         &         &         &         &         &         &       \\
5927.786  & 4.652  & --1.243   &  65.7   &  70.5   &  69.6   &  66.1   &  57.7   &  63.5   &         &  64.1   &  87.7   &  68.4   &         &  74.3   &  76.7   &  66.7   &  58.3 \\
5929.667  & 4.549  & --1.332   &  69.5   &         &         &  69.9   &  62.8   &  66.3   &  86.4   &  67.2   &         &  73.2   &  82.6   &         &         &  65.5   &       \\
5930.173  & 4.652  & --0.347   & 112.7   & 116.0   &         & 122.9   & 112.3   & 106.4   & 134.8   & 117.0   & 160.6   & 126.5   & 127.1   & 137.8   &         & 118.9   &       \\
5947.503  & 4.607  & --2.059   &         &         &         &         &         &  23.1   &  46.1   &         &         &  28.5   &         &         &         &         &       \\
6078.491  & 4.796  & --0.414   &  97.5   & 104.3   & 106.8   & 112.3   &  93.7   &  97.1   & 121.9   & 104.0   & 126.1   & 110.8   & 110.6   & 119.8   & 115.6   & 104.2   &  87.8 \\
6078.999  & 4.652  & --1.123   &  71.8   &  73.3   &  75.2   &  77.1   &  66.9   &  70.3   &         &  74.0   &  95.3   &  79.4   &         &  82.8   &  83.8   &  77.1   &  65.9 \\
6094.364  & 4.652  & --1.749   &  43.6   &  43.2   &  44.7   &  50.6   &  31.1   &  35.7   &  61.0   &  38.3   &         &  47.6   &         &  49.4   &  52.9   &  39.4   &  36.8 \\
6098.280  & 4.559  & --1.940   &  41.4   &         &  43.2   &  43.4   &         &         &         &  38.5   &         &  45.6   &         &         &  51.6   &         &  29.6 \\
6151.617  & 2.176  & --3.486   &  98.6   &  98.3   &  99.2   &  89.3   & 107.1   &  85.4   & 126.6   & 104.0   & 149.6   & 114.1   & 113.7   & 115.0   & 110.8   & 107.2   &  88.0 \\
6165.361  & 4.143  & --1.645   &  77.8   &  78.1   &  78.6   &  77.3   &  76.8   &  71.9   &  96.9   &  76.2   & 119.7   &  86.3   &  86.8   &  93.3   &  89.4   &  76.1   &  68.0 \\
6187.987  & 3.944  & --1.740   &  81.1   &  83.8   &  89.6   &  82.8   &  78.5   &  76.2   & 103.9   &  83.3   & 119.6   &  93.1   &  94.3   &  98.7   &  92.0   &  80.5   &       \\\hline
\end{tabular}
\end{table*}

\addtocounter{table}{-1}
\begin{table*}
\caption{Continued.}
\hspace*{-1.3cm}
\scriptsize
\begin{tabular}{lcrrrrrrrrrrrrrrrr} \hline\hline
  $\lambda$ (\AA) & $\chi$ (eV)  & log $gf$ & \multicolumn{15}{c}{EW (m\AA)}\\
          &        &           &\object{HD 28}    & \object{HD 1227} &\object{HD 4482}  &\object{HD 17006} &\object{HD 19754} &\object{HD 154619}&\object{HD 156266}&\object{HD 181809}&\object{HD 182776}&\object{HD 202134}&\object{HD 204128}&\object{HD 205249}&\object{HD 211391}&\object{HD 217188}&\object{HD 218527}\\\hline
6219.279  & 2.198  & --2.551   & 147.0   & 149.9   & 156.1   & 142.3   & 164.9   & 132.3   &         &         &         &         &         &         & 164.6   & 166.4   & 132.9 \\
6252.554  & 2.404  & --1.867   &         &         &         &         &         & 164.0   &         &         &         &         &         &         &         &         & 160.0 \\
6322.690  & 2.588  & --2.503   & 127.8   & 125.1   & 129.9   & 120.9   & 134.3   & 112.6   & 161.5   & 130.1   &         &         & 146.9   & 153.0   & 138.5   & 135.9   & 110.2 \\
6335.328  & 2.198  & --2.432   & 159.0   & 152.9   & 166.6   & 151.8   & 173.2   & 141.7   &         &         &         &         &         &         & 171.4   & 171.1   & 137.3 \\
6336.823  & 3.687  & --0.896   & 141.8   & 139.4   & 150.4   & 154.5   & 152.6   &         & 169.7   &         &         &         & 171.1   &         & 152.1   & 160.3   &       \\
6436.411  & 4.187  & --2.538   &  36.7   &  37.9   &         &  35.2   &         &         &         &  27.3   &         &         &         &         &         &         &       \\
6593.871  & 2.433  & --2.342   &         &         &         &         & 159.6   &         &         &         &         & 173.8   &         &         &         &         & 125.1 \\
6699.162  & 4.593  & --2.172   &  27.6   &         &         &  31.3   &         &  26.1   &  45.5   &  22.3   &         &  25.3   &         &  30.4   &         &         &       \\
6713.771  & 4.796  & --1.606   &  46.7   &  46.3   &  47.4   &  52.9   &  36.4   &         &  65.5   &  41.9   &         &  49.3   &         &  59.0   &  55.7   &         &       \\
6725.353  & 4.104  & --2.370   &  43.1   &  45.7   &  47.9   &  44.5   &         &  38.3   &  63.4   &  39.5   &         &  43.9   &         &  49.1   &  51.7   &  43.2   &       \\
6726.661  & 4.607  & --1.200   &  70.7   &  78.8   &  75.5   &  76.3   &  70.9   &  67.5   &  91.1   &  72.0   &         &  76.1   &  80.0   &  83.8   &  84.5   &  77.8   &  61.6 \\
6733.151  & 4.639  & --1.594   &  52.6   &  58.0   &  54.5   &  56.2   &  47.5   &  51.5   &  72.8   &  48.1   &         &  57.1   &  59.3   &  61.7   &  65.2   &  50.2   &  41.7 \\
6745.090  & 4.580  & --2.192   &  23.6   &         &  27.6   &         &         &  21.7   &         &  24.3   &         &         &         &         &         &         &       \\
6750.150  & 2.424  & --2.727   & 124.1   & 127.7   & 128.0   & 120.3   & 135.4   & 114.6   & 160.2   & 133.7   &         & 146.5   & 151.9   & 151.4   & 138.3   & 142.6   &       \\
6806.847  & 2.728  & --3.265   &  84.8   &         &  82.7   &  75.4   &  85.0   &  73.3   &         &  83.5   & 129.7   &  99.1   &         &         &  91.1   &  91.5   &  75.7 \\
6810.257  & 4.607  & --1.129   &  73.3   &  83.6   &  81.3   &  85.9   &  74.5   &  74.8   &  98.8   &  79.7   & 117.8   &  84.7   &  94.5   & 100.0   &  90.2   &  82.1   &  74.4 \\
6820.369  & 4.639  & --1.289   &  71.3   &  76.1   &  76.1   &  76.0   &  60.4   &  68.3   &         &  68.8   &         &  78.7   &         &  80.9   &  81.4   &  71.5   &  63.8 \\
6843.648  & 4.549  & --0.934   &  86.2   &  95.3   &  94.6   &  97.2   &  82.8   &  86.4   & 111.8   &  91.4   &         &  94.5   &  94.5   & 103.0   & 105.4   &  88.7   &  77.4 \\
6857.243  & 4.076  & --2.203   &  51.7   &  53.2   &  53.1   &  50.2   &         &  48.7   &  73.4   &  47.2   &         &         &  53.8   &  57.5   &  64.0   &  48.2   &  42.2 \\
6862.492  & 4.559  & --1.509   &  56.4   &  62.8   &  61.5   &  64.1   &  49.2   &  62.5   &  80.8   &         &         &  64.7   &  64.2   &         &  75.3   &  54.4   &       \\
6988.523  & 2.404  & --3.623   &         &         &         &         &         &  74.0   &         &         &         &         &         &  97.7   &         &  83.8   &  72.1 \\
7022.953  & 4.191  & --1.184   &  92.9   &         &  99.0   &         &         &         &         &         & 133.4   & 104.4   &         & 113.1   &         & 100.2   &       \\
7219.678  & 4.076  & --1.715   &         &         &         &  78.9   &         &  73.3   &  97.5   &         &         &  84.0   &         &         &         &         &       \\
7306.556  & 4.178  & --1.684   &         &         &         &  78.9   &         &         & 100.2   &         &         &  86.6   &  84.8   &         &         &         &       \\
7746.587  & 5.064  & --1.379   &  43.3   &  47.5   &         &  50.8   &         &         &         &         &         &  46.1   &         &         &  52.4   &         &  40.0 \\
7748.274  & 2.949  & --1.748   & 163.5   & 162.7   & 170.2   & 161.3   & 164.3   & 147.1   &         & 168.4   &         &         & 187.8   & 189.8   & 180.5   & 169.8   & 146.1 \\
7751.137  & 4.992  & --0.841   &  77.2   &  82.7   &         &  82.0   &  70.7   &  75.7   &         &  71.3   &         &  82.8   &  91.9   &  96.8   &         &  77.2   &  68.6 \\
7780.552  & 4.474  & --0.175   & 142.9   & 148.4   & 152.2   & 160.3   & 149.1   & 140.6   & 168.9   & 155.8   & 193.0   & 162.1   & 166.4   & 173.3   & 159.1   & 154.8   & 137.7 \\
7802.473  & 5.086  & --1.493   &  35.9   &  37.2   &  38.4   &  39.4   &         &  33.3   &         &         &         &  36.1   &         &         &         &         &  30.1 \\
7807.952  & 4.992  & --0.602   &  88.6   &  98.8   &  93.2   &  95.5   &         &  85.4   & 105.5   &  86.4   & 128.1   &  94.8   & 103.0   & 105.0   & 104.4   &  84.3   &  76.8 \\
8239.127  & 2.424  & --3.433   &         &         &         &  91.3   & 103.8   &  82.5   &         &         &         & 117.4   &         &         & 110.9   & 102.4   &       \\
8592.945  & 4.956  & --0.567   &         &         &         & 102.9   &         &  93.0   & 114.5   &         &         &         &         &         & 105.5   &         &       \\
{\bf \ion{Fe}{ii}}; $\log$ $\epsilon_{\odot}$=7.67&&&&&&&&&&&&&&&&\\
5991.376  & 3.153  & --3.702   &  46.7   &         &         &  54.0   &         &  52.7   &  58.2   &  41.2   &  61.2   &  46.3   &  46.8   &  55.4   &         &  44.9   &  50.8 \\
6149.258  & 3.889  & --2.858   &  38.5   &  60.9   &  54.7   &  50.2   &  41.8   &  54.0   &  53.5   &  45.5   &         &  46.2   &         &  61.9   &  63.6   &  50.8   &  55.5 \\
6369.462  & 2.891  & --4.192   &  30.6   &  41.1   &         &  36.4   &         &  38.3   &         &         &         &         &         &         &  49.9   &  35.6   &       \\
6416.919  & 3.892  & --2.750   &  47.4   &  63.8   &  59.4   &  57.4   &  50.2   &  58.1   &         &  51.8   &  66.1   &  54.9   &  59.2   &  63.6   &  66.4   &  52.7   &  52.1 \\
6456.383  & 3.904  & --2.209   &  64.2   &  84.4   &  86.7   &  78.9   &  78.3   &  79.4   &  76.3   &  70.8   &  88.3   &  75.8   &  74.5   &  94.4   &  93.1   &  77.8   &  78.0 \\
7711.723  & 3.904  & --2.625   &         &         &         &  66.7   &  48.7   &  67.4   &  62.1   &  52.4   &         &  56.5   &         &         &         &  58.7   &       \\
{\bf \ion{Co}{i}}; $\log$ $\epsilon_{\odot}$=4.92&&&&&&&&&&&&&&&&\\
6454.990  & 3.632  & --0.233   &         &  46.9   &         &  50.2   &  38.3   &  37.5   &         &  40.8   &         &  52.7   &  56.1   &  60.6   &         &         &       \\
{\bf \ion{Ni}{i}}; $\log$ $\epsilon_{\odot}$=6.25&&&&&&&&&&&&&&&&\\
5593.733  & 3.899  & --0.683   &  74.3   &  71.9   &  70.1   &  80.5   &  59.9   &  64.9   &  90.7   &  65.5   &  77.6   &  74.4   &  81.0   &  85.9   &  80.9   &  63.0   &  63.0 \\
5805.213  & 4.168  & --0.530   &  61.9   &  68.8   &  64.5   &  73.3   &  55.1   &  59.5   &  83.9   &  61.6   &  81.9   &  67.8   &         &  74.6   &  73.8   &  60.9   &  57.2 \\
6111.066  & 4.088  & --0.785   &  60.7   &  63.7   &  61.3   &  68.2   &  47.4   &  53.7   &  77.6   &  55.6   &  80.6   &  63.5   &  65.0   &  70.9   &  71.0   &  55.7   &  50.6 \\
6176.807  & 4.088  & --0.148   &  89.6   &  97.7   &  94.6   &  98.3   &  84.6   &  85.7   & 111.7   &  88.1   &         &  98.6   &  96.8   & 108.9   & 105.6   &  86.8   &  79.4 \\
6186.709  & 4.106  & --0.777   &  54.7   &  62.9   &  63.7   &  67.6   &         &  53.2   &  76.5   &  54.9   &         &  68.5   &  62.5   &  69.0   &  72.2   &  54.5   &  47.8 \\
6204.600  & 4.088  & --1.060   &  48.9   &  52.9   &  51.4   &  54.9   &  36.5   &  42.0   &  72.9   &  39.7   &         &  51.3   &  46.0   &  65.1   &  62.7   &         &  40.1 \\
6223.981  & 4.106  & --0.876   &  52.4   &  57.2   &  58.7   &  61.9   &         &  51.8   &  78.9   &  52.0   &         &  62.9   &  59.7   &  70.5   &  67.1   &         &       \\
6772.313  & 3.658  & --0.890   &  74.4   &  81.8   &  76.2   &  80.8   &  75.8   &  69.6   & 102.7   &  77.1   & 116.4   &  86.8   &         &  99.5   &  87.4   &  77.2   &       \\
7555.598  & 3.848  &   0.069   & 127.1   & 132.0   & 130.1   & 137.9   & 120.7   & 115.5   & 157.8   & 126.1   & 160.9   & 135.7   & 140.4   & 153.1   & 144.6   & 126.8   & 114.8 \\
7797.586  & 3.899  & --0.144   & 106.6   & 111.3   & 113.9   & 112.4   & 107.4   & 102.3   & 125.6   & 108.4   &         & 113.6   & 118.4   & 122.6   & 123.6   & 108.3   & 100.0 \\
{\bf \ion{Ba}{ii}}; $\log$ $\epsilon_{\odot}$=2.13&&&&&&&&&&&&&&&&\\
5853.668  & 0.604  & --0.758   & 106.7   & 121.2   & 126.7   &  91.1   & 144.6   & 116.2   & 129.1   & 127.4   & 162.3   & 134.1   & 132.2   & 133.9   & 131.2   & 137.6   & 110.2 \\\hline
\end{tabular}
\end{table*}

\section{NLTE corrections} \label{sect_b1}
Table~\ref{tab_b1} gives the NLTE corrections applied to the abundances of Mg, Na and Li. For completeness, we also include the stars analyzed in Paper II.

\begin{table*}
\centering
\caption{NLTE corrections applied to the abundances of Mg, Na and Li (Gratton \etal 1999; Carlsson \etal 1994). The corrections are defined as: $\Delta \epsilon=\log(\epsilon)_{\rm NLTE}-\log(\epsilon)_{\rm LTE}$.}
\label{tab_b1}
\begin{tabular}{lcccc} \hline\hline
Star               & \ion{Mg}{i} $\lambda$5711 & \ion{Na}{i} $\lambda$6154 & \ion{Li}{i} $\lambda$6707\\\hline 
\object{HD 28    } & +0.06 & +0.02 & +0.22\\
\object{HD 1227  } & +0.06 & +0.02 & +0.15\\
\object{HD 4482  } & +0.06 & +0.00 & +0.17\\
\object{HD 10909 } & +0.05 & --0.02 & +0.20 \\
\object{HD 17006 } & +0.08 & --0.04 & +0.10\\
\object{HD 19754 } & +0.03 & --0.09 & +0.24\\
\object{HD 72688 } & +0.06 & --0.03 & +0.15 \\
\object{HD 83442 } & +0.06 & --0.06 & +0.24\\
\object{HD 113816} & +0.06 & --0.03 & +0.24\\
\object{HD 118238} &       & --0.09 & +0.28\\
\object{HD 119285} & +0.06 & --0.03 & +0.22\\
\object{HD 154619} & +0.10 & --0.09 & +0.12\\
\object{HD 156266} & +0.04 & --0.06 & +0.26\\
\object{HD 181809} & +0.06 & +0.02 & +0.19\\
\object{HD 182776} &       & --0.09 & +0.25\\
\object{HD 202134} & +0.05 & --0.03 & +0.23\\
\object{HD 204128} & +0.06 & --0.04 & +0.20\\
\object{HD 205249} & +0.06 & --0.06 & +0.13\\
\object{HD 211391} & +0.06 & --0.03 & +0.16\\
\object{HD 217188} & +0.06 & --0.01 & +0.20\\
\object{HD 218527} & +0.06 & +0.03 & +0.15\\\hline 
\end{tabular}
\end{table*}

\end{document}